\documentclass[galaxies,review,accept,pdftex,oneauthor]{Definitions/mdpi}
\usepackage{upgreek}
\usepackage[none]{hyphenat}

\firstpage{1}
\pubvolume{11}
\issuenum{3}
\articlenumber{74}
\pubyear{2023}
\copyrightyear{2023}
\externaleditor{Academic Editors: { Thomas W. Jones and Stanislav S. Shabala } 
}
\datereceived{29 March 2023}
\daterevised{25 May 2023} 
\dateaccepted{26 May 2023}
\datepublished{13 June 2023 }
\hreflink{https://doi.org/\linebreak10.3390/galaxies11030074} 

\def\pasa{Pubs.\ Astron.\ Soc.\ Australia}

\Title{\textls[-15]{The Dynamics and Energetics of Remnant and Restarting~RLAGN}}

\TitleCitation{The Dynamics and Energetics of Remnant and Restarting RLAGN}

\Author{Vijay 
{H.} 
Mahatma \href{https://orcid.org/0000-0001-5221-2636}{\orcidicon}}

\AuthorNames{Vijay H. Mahatma}

\AuthorCitation{Mahatma, 
V.H.}

\address[1]{%
Thuringer Landessternwarte, Sternwarte 5, D-07778 Tautenburg, Germany; {v.mahatma93@gmail.com} 
}

\abstract{In this article, I review past, current, and future advances on the study of radio-loud AGN (RLAGN; radio-loud quasars and radio galaxies) lifecycles exclusively in the remnant and restarting phases. I focus on their dynamics and energetics as inferred from radio observations while discussing their radiative lifetimes, population statistics, and trends in their physical characteristics. I briefly summarise multi-wavelength observations, particularly X-rays, that have enabled studies of the large-scale environments of RLAGN in order to understand their role in feedback. Furthermore, I discuss analytic and numerical simulations that predict key properties of remnant and restarting sources as found in wide-area surveys, and discuss the prospects of future surveys that may shed further light on these elusive subpopulations of RLAGN. }

\keyword{active galactic nuclei; jets; radio observations; radio galaxy lifecycles; simulations}

\begin{document}
\section{Introduction}
Radio-loud 
active galactic nuclei (RLAGN) are a class of active galaxies that contain an accreting supermassive black hole at the centre (called AGN). RLAGN objects are distinguished from AGN in general by their observed emissions at radio wavelengths. AGN (or radio-quiet 
AGN in this context) tend to have little or no radio emission measurable over that which describes star formation in their host galaxies. On the other hand, RLAGN tend to have bright radio emission that is driven by bipolar jets (thought to be produced by the supermassive black hole and its accretion system containing strong magnetic fields, e.g., \cite{blan77,bege84,blan19}). The radio emission as a whole (seen on pc to Mpc scales) is produced by the synchrotron process, where relativistic charged particles gyrate around magnetic fields; this can be seen in other wavebands as well (in fact, the very first detection of an astrophysical jet associated with RLAGN was made in the optical band \citep{curt18}). Detailed studies of RLAGN using radio observations at very high angular resolution to understand their microphysics have shed light on their intrinsic structure, termination points, and the associated particle acceleration (e.g., \cite{lain80,blac92,hard97,fern97,mull06}), as have multi-wavelength observations on the description of their overall impact on the evolution of galaxies (see the review by \cite{fabi12}).

In the past, the distinction between RLAGN and AGN was clear; traditionally, a `radio-loudness' parameter $R$ separated the two classes, defined as $R=f(4400\,\text{\AA})/f(6\,\text{cm})$, i.e., the optical to radio total flux \citep{kell89}, with RLAGN having $R\gtrsim10^3$ and radio-quiet AGN having $R\lesssim10$ (predominantly coming from star-formation \citep{wils95}) and lacking evidence of large-scale jets. This is equivalent to the implication that, for RLAGN, a large fraction of their total bolometric luminosity is in the form of the kinetic power of the jets. Distinctions have been made in the X-ray regime as well, where contamination from galaxy starlight is not an issue, such as $R_X=\nu L_{\nu}\,(6\,\text{cm})/L (2\text{--}10\,\text{keV})$ \citep{tera03}. However, such sharp distinctions are hard to make, as there is now robust evidence for radio-quiet AGN having jets and lobes (e.g., \cite{yao21}), while RLAGN are being probed to lower radio powers than ever before (e.g., \cite{ming19}). Rather, there is likely a continuous distribution of jet power across the AGN population, with the higher jet power population more likely to be hosted by elliptical galaxies (perhaps favouring accretion through a strong merger history and higher black hole spin \citep{wils95}). Nevertheless, distinguishing between RLAGN and radio-quiet AGN is important in order to select samples that can help to understand the nature and origin of their radio emissions. In this review, I make the distinction that RLAGN are those AGN that are selected by their morphology at radio frequencies (i.e., through the clear presence of jets, lobes, and/or hotspots), while radio-quiet AGN are those AGN that are selected by every other wavelength regime (i.e., optical and X-ray spectra, infrared colours, etcetera). It is particularly important for the purposes of this review that RLAGN samples are best suited to understand the lifecycles of jet activity amongst the AGN population.

The idea that RLAGN (I 
use this term to include both radio galaxies and radio-loud quasars, but not blazars or BL Lac objects, which are strongly beamed and are not discussed in this review) have episodic activity, meaning that jet production is intermittent, recurrent, or simply has a finite lifetime, has been suggested in many forms, including deep high-\linebreak resolution radio imaging, radio source counts, and multi-wavelength studies. Those RLAGN with recently switched-off or restarting jets are an important subpopulation, as by definition they provide insights into lifecycles as well as the jet-triggering and feedback processes. Owing to recent advances in the study of RLAGN lifecycles, here I review the literature on remnant and restarting RLAGN from an observational perspective, in radio and other wavebands and from the perspective of models and simulations of their radio emission.

I first point out, as a note of caution with respect to nomenclature, that \textit{intermittent} generally refers to the situation of stopping and starting of jets at rapid intervals, sometimes indefinitely, whereas \textit{recurrent} and \textit{restarting} activity is generally used to suggest longer (Myr) timescales between episodes. For the purposes of this review, I use the terms interchangeably, as the words have similar meaning and have been used interchangeably in the literature. Intermittency has been used to describe the ejection of discrete electron populations up the jet on very short (tens or hundreds of yr) timescales, however, this is not the focus of the present review. I additionally note that in the past the term “relic” has been used to describe a source which is no longer actively producing jets; however, this terminology has recently become deprecated, as this term is more often used to describe bright arc-like structures at the peripheries of galaxy clusters associated with cluster shocks (which, ironically, may originate from remnant RLAGN plasma). I therefore use the term “remnant” exclusively to describe a radio emission from RLAGN that does not have a detectable jet. For information on RLAGN in general, I refer the reader to reviews such as~\cite{hard20}.

What is the current picture of the various life-cycles of RLAGN, how complete is it, and how important are the remnant and restarting phases to galaxy evolution? First, the important distinctions that allow us to distinguish the remnant and restarting RLAGN subpopulations from the general population are their (predominantly radio) morphology, spectral energy distribution, and disturbances in their hot gas media unassociated with \textit{current} radio emission. RLAGN begin their lifecycle as pc-scale jets, observed as compact sources ($\leq$10 kpc; \citep{odea98}), where newly-born jets drive through the dense interstellar medium \textls[-15]{of their host galaxies. Surprisingly, and as I detail below, there is a significant over-abundance} of these sources, suggestive of the view that radio jets terminate prematurely. RLAGN can reach scales of $\gtrsim$100 kpc, where they can generally be distinguished morphologically into Fanaroff--Riley type I and II sources (FR-I and FR-II \citep{FR74}), although there are a variety of sources that do not fit into this distinction. The morphological and energetic evolution of these large-scale jet-driven sources are widely believed, in their later stages, to describe a cessation or dramatic drop in jet activity. Jets are known to restart after a period of quiescence, and many sources restart with a similar jet power to the previous episode of activity. It is not known whether all RLAGN go through all or even some of these evolutionary paths, or whether these paths are restricted to any classes of RLAGN. Most importantly, it is not clear what causes RLAGN to cease radio activity over any timescale, as certain objects reach Mpc scales and others apparently terminate their activity on scales that are orders of magnitude lower. Because RLAGN in general are important in models of galaxy evolution, their disrupted activity and how this affects their ability to affect their surroundings is important to understand in detail. Remnant radio galaxies that do not currently have an active jet may be important for the suite of models describing AGN feedback, as the total energy supplied by the jet during its lifetime remains stored in the lobes even if these have faded beyond the point of being detectable with radio telescopes.  The time and spatial scales on which this energy reservoir impacts the surrounding environment, as well as the resulting consequences for galaxy evolution, remain open questions.

Numerical simulations and (semi-)analytic models that capture the key physics of radio lobe plasma have progressed in recent years, offering new insights into the detectability and evolution of remnant and restarting sources. Because observational advances must drive a better theoretical understanding of these rare subclasses of RLAGN, I focus more on the observational progress while highlighting the important theoretical results. I particularly emphasize results from recent low-frequency radio observations, summarizing them in relation to current outstanding questions and to those that may be answered with upcoming radio instruments. Multi-wavelength data (i.e., on wavelengths shorter than radio) are important as well, offering insights into RLAGN energetics and their environments that cannot be probed in the radio regime. I would point interested readers to past reviews on this subject that may focus in more detail on certain aspects of RLAGN lifecycles than this review can \citep{sari12_review,morg19}.

In Section \ref{sect:obs}, I review observational studies of remnant and restarting RLAGN, provide a historical overview, then move to an account of recent low-frequency and multi-wavelength studies. In Section \ref{sect:simulations}, I discuss the key predictions made by models and simulations of remnant and restarted jet activity. I conclude with a summary, discuss open questions, and provide a brief outlook on studies using future next-generation instruments in Section \ref{sect:summary}. I define the spectral index in the sense of $S\propto\nu^{-\alpha}$ except as otherwise stated.
\section{Observations}
\label{sect:obs}
\subsection{Radio}
\label{sect:radio}
\subsubsection{Historical Overview of Remnant RLAGN}
\label{sect:remnants}
If jet activity has ceased completely, the hotspots are no longer being fed and the radio emission in the lobes evolves without an energy supply. This has three major implications for identifying remnant sources through radio observations: if hotspots are the visible manifestation of accelerated particles driven by the jet-environment interaction, then remnants should not contain jets, jet knots or hotspots in the lobes (except in the case where the jet has switched off but the source is observed during the light-travel time of the jet material to the hotspots). Second, if a significant amount of time has passed from cessation of activity, the large-scale lobe emission should be relatively low surface brightness over the entire source, and should become amorphous with time as the lobes expand in all directions equally. Third, a particularly long time after cessation of activity, the lobe material as a whole should be mostly composed of steep-spectrum material, and in general remnant sources should have steeper integrated spectra than active sources, as the latter are thought to have low frequency spectral indices of $\gtrsim$0.5; as such remnants must have values significantly steeper than this. The `relaxed' structure, i.e., with a low axial ratio, predicted for remnants (such sources have been associated with remnants in the giant radio galaxy catalogue by \cite{ande21}, though without re-analysis of their spectral indices or deep inspection of the lack of radio cores, which are more robust indicators of a lack of jets) is due to a lack of forward momentum and expansion along the jet axis; However, a lack of information on the timescales over which the lobe structure changes and on the axial ratio of the lobes in their previously active state makes this an inadequate criterion. For example, while many sources characterized as active FR-I sources have amorphous lobes and plumes, these have active jets. The easiest way to identify a source without jet activity is through the lack of compact flat-spectrum emission (radio cores or hotspots) at high frequencies, particularly as high frequency instruments (>1 GHz) are more sensitive than at lower frequencies, and generally have adequate resolution to distinguish between compact and diffuse material. Often, however, radio surveys are performed at lower resolution than is generally required for this distinction, and in these cases it is efficient to select \textit{candidate} remnants that have no core and steep spectral indices, then follow up with high-resolution and multi-radio-frequency data for confirmation of both. I provide a brief (likely incomplete) historical overview of such studies below.

The identification of remnant RLAGN was first made through sensitive high-frequency radio observations of their cores; strong radio emission from the nucleus of an active galaxy is clear evidence of a central engine that depicts (parsec-scale) jet activity. Therefore, a lack of radio emission at the centre of a host galaxy provides evidence suggesting that a radio galaxy has terminated jet activity. This has led to a number of studies that have performed statistical tests, mainly on the relationship between the core luminosity and the brightness of the extended lobe emission of bright objects (e.g., \cite{fere84}). In \cite{giov88}, the authors studied B2 and 3C objects with a lack of core emission using a sample from \cite{fere84}, then, through a sensitive 5~GHz follow-up, were able to find cores in 24 of 33 objects. Here, the angular resolution ($\sim$0.3 arcsec) provides adequate distinction between the core and the surrounding low surface brightness lobe material, as well as high sensitivity to faint cores. Only 8 of 187 objects in their extended sample had a lack of core detection, along with a few sources with a core that showed a relaxed structure with steep spectral indices. This 4\% remnant fraction would indicate a short remnant lifetime relative to the average synchrotron lifetime of active sources. Moreover, only 7\% of objects selected from the 3CRR catalogue \citep{3crr} at $z\leq1.0$ lacked a core detection~\citep{mull08}, implying rapid fading times if they are genuine remnants. On the other hand, it is possible that core emission is unrelated to large-scale radio emission driven by an active jet. In \cite{cord86}, the author investigated this by comparing the core powers and total powers (i.e., core luminosity vs. total luminosity of the extended emission) of objects under the assumption that a positive correlation exists if the extended emission is connected to the central engine, finding that objects beneath the known correlation show evidence of jets and hotspots. In fact, there are many examples of a tightening of the core power--total power correlation when non-detections (i.e., possible remnants) are taken into account, suggesting that secular processes such as star formation do not contribute to false positive remnant classifications, particularly when the radio core is very weak \citep[][]{burn82,burn84,fabb84}.

Steep spectra (e.g., $\alpha\geqslant$1), which are characteristic of electron populations that are undergoing ageing only or have radiated away much of their initial kinetic energy \citep{pach70}, are strong indications of RLAGN lobes not being powered by a jet. Many early radio surveys or targeted observations of a sample of objects which involved the compilation of integrated spectra provided early indications of the presence of such sources, e.g., \cite{ande88,giov91,greg92}. It should be noted that there are many `steep-spectrum' objects that do not have evidence for cores, jets, or hotspots, but lack imaging at adequate resolution and sensitivity to detect them as in the aforementioned studies. Nevertheless, there are steep-spectrum sources that do have clear evidence for jet activity, and as such a single spectral index criterion will clearly contaminate remnant samples with active sources. The inadequate resolution of past wide-area surveys (i.e., insufficient to detect cores or only resolved into a double structure) prompted higher resolution observations of sources with particularly steep spectra, as was performed for the source IC\,2476, classified as a remnant \citep{cord87} through the absence of compact structure on arc-second scales and a mean spectral index in the lobes of $\alpha\geqslant1$. A similar study was performed for the radio source 1401-33, a head-tail source suggested to be confined by cluster gas \citep{goss87}. Clusters have traditionally been known to host remnant sources due to a tendency for hosting steep spectrum sources ($\alpha\geqslant1$), first in the Coma cluster~\citep{vale78} and later in other rich Abell clusters (e.g., Abell 2255 \cite{fere97}). This can be explained by the dense hot gas in the ICM confining the radio plasma, reducing the lobe expansion speed and adiabatic losses and thereby increasing the radiative lifetime \citep{gold94} compared to a sparse field environment in which lobe material can freely expand. A particularly extreme example is the tailed radio source at the peripheries of Abell\,4038, suggested to be the remnant of an FR-II radio galaxy, with a spectral index of $\alpha=2.2$ between 80\,MHz and 1400\,MHz \citep{slee98}. On the other hand, in-falling (head-)tailed radio sources are commonly found in clusters, which are commonly steep-spectrum objects but are not necessarily remnants.

\textls[-18]{The source B2\,0924 + 30 is another well studied source considered to be a prototypical and genuine remnant; it has steep spectral indices with $\alpha\geqslant1.0$ at frequencies below \linebreak 1~GHz~\citep{eker81,cord87}, an upper-limit core luminosity  five orders of magnitude below that of the strongest core-dominated RLAGN and which is two orders of magnitude below the core luminosity expected from the core-total luminosity relation for elliptical galaxies (e.g., \cite{giov88}). With broad-band spectral aging analysis using maps at an angular resolution of 147 arcsec~\cite{jamr04}, the source has a pronounced integrated spectral steepening beyond 2\,GHz, an average spectral age of $50^{+12}_{-11}$\,Myr (since last particle acceleration, consistent with the more robust modelling by \cite{turn18}), and a maximum spectral age of the particle population of $90^{+30}_{-30}$\,Myr, similar to the derived spectral ages of other remnant lobes (e.g., B0917 + 75, 100\,Myr; \citep{harr93}, J1324-3138, 97\,Myr;~\citep{vent98}, CL\,0838 + 1948, 110\,Myr; \citep{giac21}). }

It is important to discuss the many caveats in the determination of spectral age estimates through radiative model fitting to observed spectra. One difficulty arises for genuine remnant sources that have intrinsically high spectral curvature at GHz frequencies. The model described by \cite{kard62,pach70} (KP model) can produce an exponential high frequency cutoff in the radio spectrum under the assumption of no pitch-angle scattering of the relativistic particles (and as such no isotropization, with which active lobes may be modelled), although modifications to this in order to reflect the steep-spectrum nature of remnants were presented by \cite{komi94}. I refer the reader to studies of the comparison between models for synchrotron and inverse-Compton losses (e.g., \cite{harwood13,harwood15}), and in particular to that which compares resolved and integrated models and their application to remnant sources \citep{harw17_specmodels}. Another significant caveat is that the lobe magnetic field is unknown, except for those in a few FR-II sources (e.g., \cite{cros05}) where lobe inverse-Compton emission is detected and used to determine the best-fit magnetic field strength to the radio to X-ray spectral energy distribution \citep{hard98}. The aforementioned age estimates, bearing in mind large fitting errors, must be considered as lower limits due to the use of equipartition to derive the lobe magnetic field strength. Moreover, the assumption and use of a single magnetic field strength is clearly incorrect; even a single particle population (which is never resolved observationally) experiences an evolving magnetic field strength through its lifetime, and even measured estimates using inverse-Compton emission are only instantaneous field strengths. In \cite{maha20}, the authors incorporated a time-integrated magnetic field strength for a highly resolved lobe region, finding that such a magnetic field strength can be a factor of $\sim$2 greater than the instantaneous (inverse-Compton-based) value, making the spectral age younger (and more incompatible with its dynamical age). While the magnetic field microphysics of remnant lobes are difficult to understand observationally, it is important to obtain high resolution in order to resolve lobe substructures which may have varying ages without incurring too significant a loss to sensitivity in the oldest material, as the latter provides information on the time from cessation of particle acceleration. To my knowledge, although inverse-Compton emission is largely detected from low-luminosity FR-II lobes, no such detection has been made from remnant sources. Energy equipartition between the lobe energy densities in the magnetic field and the relativistic particles has been invoked as an estimate of the magnetic field strengths (e.g., \cite{jamr04}) for both remnant and active source studies; however, this has been used for ease rather than on any physical basis, as it is well known that the magnetic field strengths measured with inverse-Compton detections in radio galaxy lobes can in general be below the equipartition value by as much as a factor of ten \citep{cros05,maha20}.

\subsubsection{Historical Overview of Restarting RLAGN}
\label{sect:restarting}
RLAGN are suggested to be restarting if there is evidence for both remnant plasma and active jets in the same object. The suggestion of restarting activity permeated early studies of radio sources, even before the introduction of remnants. The authors of \cite{burb65} speculated intermittency on Myr timescales in the famous source 3C\,84 using optical velocity dispersion measurements, while \cite{kell66} inferred radio source intermittency timescales of $\sim$$10^4$--$10^6$\,yr based on the radio spectral index distribution of the brightest radio sources. Statistical evidence for multiple outbursts was suggested to be embedded in the number counts vs. size for small luminous RLAGN, with \cite{odea97} noting a plateau in the decline of counts with size, which can be explained if young RLAGN are intermittent on time scales of $\sim$$10^5$--$10^6$\,yr. If the range of inferred intermittency timescales is robust, then the disruption timescale for a given jetted AGN is not fixed, and is perhaps set by stochastic processes (as I describe below). Notwithstanding this, these early studies drove the need for more direct timescale estimates by probing radio activity.

Because the detection of remnant plasma is a necessary prerequisite, it might be expected that restarting sources would be as easy to detect as remnants. However, the situation is complicated by the fact that the restarting jets actually drive through the remnant plasma left behind by the previous outburst. This means that the external environment of the restarting jets is not that of a typical non-restarted source (i.e., a hot gas IGM), and as such the radiative evolution of the radio source as a whole deviates from expectations. Observationally, this requires extra information or selection criteria beyond those that define remnants. For example, if remnants are defined  on the basis of not having a visible radio core, then consistency dictates that all RLAGN with a visible core can be candidate restarting sources (which might be true). Furthermore, there are many objects that might appear to be restarting on morphological grounds but which require robust evidence to confirm their nature, such as sources with bright parsec-scale jets surrounded by diffuse and/or amorphous structures. Certain prototypical FR-Is even have termination points or jet knots on kpc scales, and could be classed as restarting within simpler morphological criteria. High resolution, particularly on sub-arcsecond scales, is commonly used to interpret or even detect compact features describing actively restarting jets. At the same time, short baselines of radio interferometric observations are needed in order to detect the diffuse large-scale remnant lobes at a broad range of frequencies to confirm the lack of flat-spectrum components. Unlike in the case of remnants, however, there are examples of unambiguously restarting sources that are easy to identify with radio observations alone, as I review below.

\begin{figure}[H]
\includegraphics[scale=0.3]{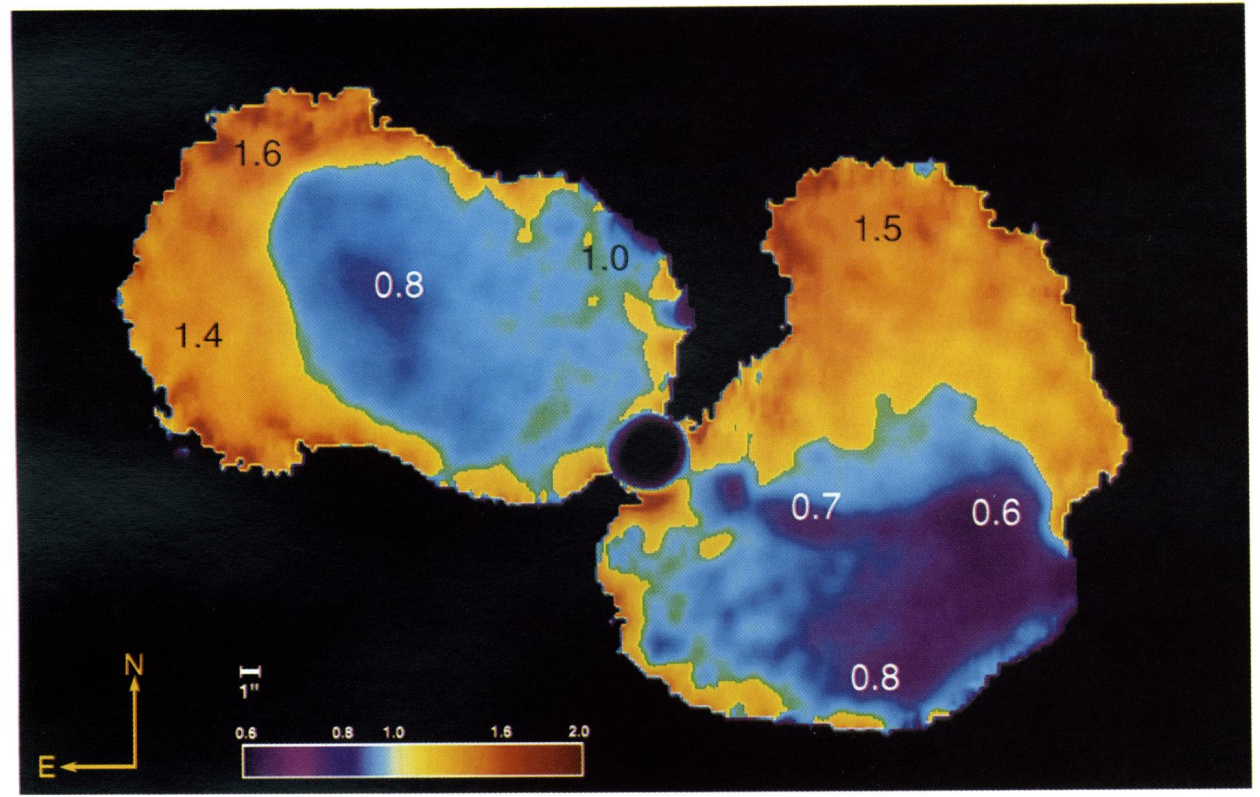}
\caption{{1.4--5} 
GHz spectral index map of 3C\,388, presented by \cite{roet94}. The sharp transition from flat ($\sim$$0.7$) to steep ($\sim$$1.5$) spectral indices can be seen between the regions coloured in blue and orange. ©~AAS. Reproduced with permission.}
\label{fig:roetigger+94}
\end{figure}

One of the earliest detailed studies of an RLAGN classed as a restarting source was of 3C\,388 \citep[][]{roet94,burn82_3c388}, a double-lobed FR-II source at the centre of a poor cluster with a jet structure terminating well within the western lobe (as opposed to towards its edge), thereby showing an asymmetrical structure on scales of a few arcsec. The most important feature when ascribing intermittency to a source with a detectable jet is the existence of remnant lobes, evident through either a lack of hotspots or steep-spectrum emission. Because particle acceleration of jet material takes place in the lobes, the lobes (in some regions) should have flatter spectral indices than the jets themselves. Thus, if the flattest spectral index in the lobes is significantly \textit{steeper} than that of the jet, then it is likely that the lobe describes remnant emission. With higher resolution and dynamic range from images of 3C\,388, Ref.~\citep[][]{roet94} detected a sharp transition in spectral index in both lobes, with $\alpha^{1400}_{5000}\approx0.8$ in the lobe region containing the hotspot and a transition discontinuous by more than 45\% to $\alpha^{1400}_{5000}\gtrsim1.4$ at a C-shaped boundary in the eastern lobe (see Figure \ref{fig:roetigger+94}). While orientation and relativistic effects can explain diffuse steep-spectrum lobe material beyond hotspots, the diffuse nature of the `hotspot' in the eastern lobe and the sharp transition to steep spectral indices in the lobe itself cannot be explained in this way. Rather, it is likely that restarted jets are driving through and interacting with aged remnant plasma from previous activity. A more dramatic example is the famous source Hercules A, hosting prominent rings within the lobes which have flatter spectra than the surrounding steep-spectrum lobe material, along with a steep-spectrum core, a structure that has been suggested to show multiple outbursts of activity \cite{giza03,timm22}.

\begin{figure}[H]
\includegraphics[scale=0.3]{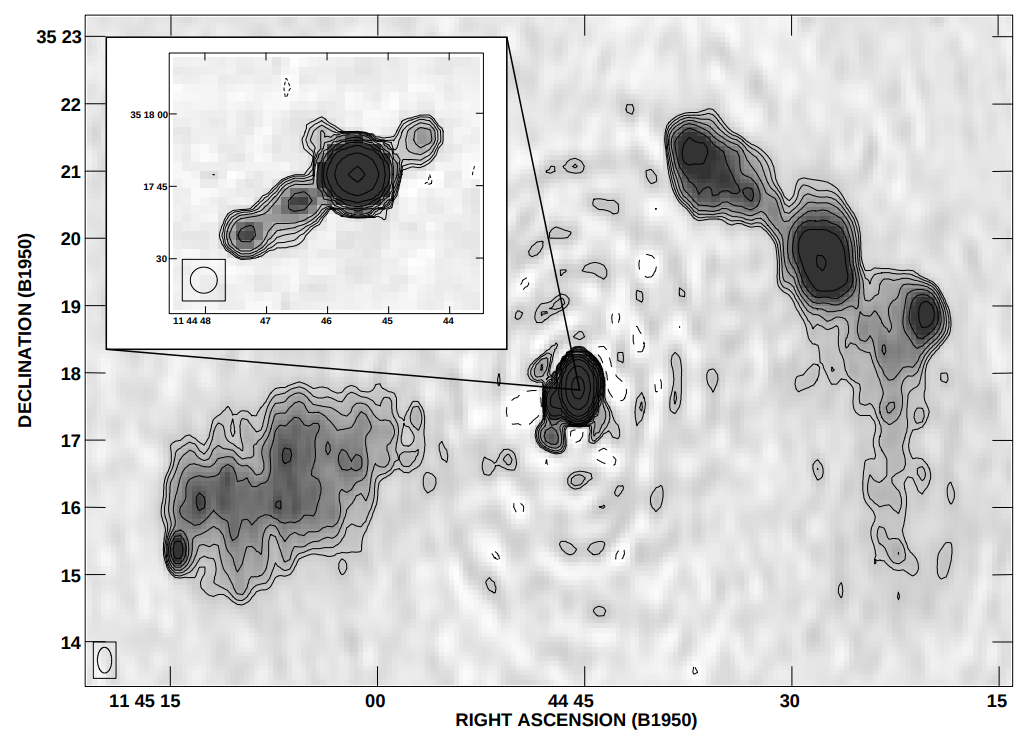}
\caption{Radio maps of the restarting source B1144 + 352 at 1.4\,GHz from \cite{scho99}. The inset shows a high resolution image of the central GPS source, showing evidence of an edge-brightened restarting jet. Reproduced with permission © ESO.}
\label{fig:schoenmakers+99_ddrg}
\end{figure}

Restarted jets are necessarily young ($\lesssim$$10^6$ yr) and generally compact, except for special cases like 3C\,388 above, making it hard to perform the same analysis for the restarting RLAGN population. There are classes of compact radio sources with jets on subgalactic scales, such as Gigahertz Peaked Spectrum (GPS; \cite{spoe85}) sources, which have inverted spectra that peak at around 1\,GHz before steepening, or Compact Steep Spectrum (CSS; e.g., \cite{fant89}) sources, which peak at lower frequencies. These compact sources are obvious candidates to search for restarted activity if they have large-scale extended lobes from a previous episode; see the review of CSS sources by \cite{odea21} and references therein. GPS sources are usually compact, even on scales probed by VLBI observations (i.e., linear sizes up to a few hundred pc~\cite{dall95,stan97}), however, many have a symmetric double radio structure as well; these are called Compact Symmetric Objects (CSO, e.g., \cite{wilk94}). While models for their origin explain these sources as being `frustrated' jets confined by a high-pressure ISM (e.g.,~\cite{vanb84,odea91}), a small percentage of these sources do indeed have extended kpc-scale emission (e.g., \cite{odea98,hanc10}). Several bright 3CRR sources have shown evidence of their bright central components belonging to the class of CSO objects, and are suggested to have restarted activity (e.g., 3C\,317 and 3C\,84 \cite{vent04,boeh93}). In particular, 3C\,84 is a prime example; it is one of the brightest and closest objects ($z=0.0176$) with small-scale restarted jets that can be tracked on human time-scales. It has been claimed that the most recent jet activity began in 2005 \citep{naga10}, extending by $\sim$1 pc from the radio core, with recently enhanced polarization from the hotspot in 2015 \citep{naga17}. The most plausible model to explain these systems is that the extended radio emission is disconnected from the young jet activity describing the compact object. Other examples of these systems are the radio sources B0108 + 388 \citep{owsi98} and B1144 + 352 (see Figure \ref{fig:schoenmakers+99_ddrg} \cite{scho99}).

One of the most important findings regarding restarting sources with pc-scale jets is that they tend to drive neutral atomic and molecular gas outflows in their surrounding ISM~\citep{morg98,morg05}. In these objects, absorption lines associated with neutral atomic hydrogen (HI) or molecular CO gas (with derived velocities of $\lesssim$$1000$\,km\,s$^{-1}$ with respect to the systemic velocity in the surroundings) are spatially aligned with the approaching or receding jet. This provides strong indications that restarted jets have a profound impact on the gas reservoirs in the ISM. This has been observed in a variety of systems, from powerful radio-luminous jets \citep{chan11,morg13,emon16} to weak and low-power jets, e.g., Seyfert galaxies \citep{oost00,morg13_seyfert,mukh18}. What is interesting is that this situation seems to occur at a higher rate in compact restarting systems than in more extended RLAGN \citep{holt08}. These cold gas outflows ($T\lesssim10^4$\,K) are most important to jet-ISM feedback, as they carry higher mass outflow rates than those in hot/warm ionized gas outflows (e.g., \cite{cico14}, and see review by \cite{veil20}). In this picture, a restarting jet imparts energy and momentum efficiently into its surroundings, driving pre-existing cold gas (e.g., HI or CO) from pc to kpc scales in the galaxy outskirts. This may lead to negative feedback, depleting gas in the inner regions of the galaxy available for star formation, or the jets can impart angular momentum to the cold gas that ought to trigger star formation. Nevertheless, the restarting jets, particularly in powerful radio-loud objects (e.g., 3C\,236, 3C\,293; \citep{schu18,schu21}), have the means to disrupt the ISM.

On slightly larger scales beyond the host galaxy ISM, several sources show evidence for a restarting jet extended along the same axis as the remnant lobes, called double-double radio galaxies (DDRGs \citep{scho99_ddrg,scho00}). These restarted jets are generally older than those in CSS or GPS sources, as the edge-brightened `inner' double is visible on kpc scales (generally required for us to resolve it, meaning that there is an observational selection effect). Among the brightest DDRGs are 3C\,445 \citep{leah97} and 4C\,26.35 \citep{owen97_ddrg}, while there are a number of \textit{candidate} DDRGs which require either high resolution observations to confirm the inner double or shorter baseline observations to confirm the outer double (e.g., 3C\,16, 3C\,424; \citep{leah91,blac92}). In the DDRG model, the new jets drive through the old low-density remnant lobes, and it might be expected that the jets do not impact with sufficiently dense material for a termination shock to develop. The detection of edge-brightened structures at the termination points of the restarting jets suggests that the density in the remnant lobes is high enough for shock-driven hotspots to form; see the discussion below on bow shocks in remnant lobes. In \cite{kais00}, it is suggested that $T\sim10^4$\,K warm gas clouds from the IGM can disperse over the remnant lobe volume, allowing for an IGM-like density such that shock-induced hotspots can form; I refer the reader to Section \ref{sect:simulations} for a further discussion of DDRG models. Nevertheless, DDRGs are the easiest of restarting sources to identify, requiring only the detection of two pairs of radio lobes along the same jet axis. They are expected to be short-lived objects, as the restarting jets are relativistic and merge with the remnant outer lobes on timescales approximately equal to the light-travel time ($t_{merge}\approx3\times10^6$\,yr for a lobe of length 100 kpc and a restarting jet travelling at $v=0.1c$), which is significantly lower than the radiative timescales of aged lobes. In Table \ref{tab:source_table} I list some noteworthy  restarting (as well as remnant) sources that are well studied in the literature.
\begin{table}[H]
\caption{Notable 
{and} 
well-studied 
remnant and restarting sources. Note that the references are selected and likely incomplete.}
\newcolumntype{C}{>{\centering\arraybackslash}X}
\begin{tabularx}{\textwidth}{CC}
\toprule
\textbf{Name} & \textbf{Reference} \\ \midrule
\textbf{Remnant} & \\
B2\,0924 + 30 & \cite{shul17} \\
Arp\,187 & \cite{ichi16}\\
J1615 + 5452 & \cite{rand20}\\
Blob1 & \cite{brie16}\\
\textbf{Restarting} & \\
3C\,84 & \cite{naga17}\\
3C\,388 & \cite{roet94,burn82_3c388,brie20} \\
4C\,29.30 & \cite{siem12}\\
3C\,236 & \cite{shul19}\\
3C\,293 & \cite{brid81,akuj96,besw02,mach16,kukr22}\\
\bottomrule
\end{tabularx}
\label{tab:source_table}
\end{table}

What if a restarted jet does not drive along the same axis as the previous activity? If there is a disruption of the accretion mechanism, it is plausible that the restarted jet can drive along a significantly different axis from the previous episode. One explanation for the morphology of X-shaped sources, for example, is an abrupt change in jet axis (note that there is no general agreement on whether the jet simply changes direction abruptly or switches off and restarts in a different direction, though deflected backflow models seem favoured; see \citep{cape02,gopa12}). Thus, they may not necessarily be restarted systems, in the sense that the current jet activity producing radio emission occurs at a large angle from the previous episode (e.g., \citep{leah84}). Alternatively, there is evidence that, rather than being restarted or jets with deflected flows, there may be multiple jets driven by multiple RLAGN in proximity (e.g.,~\citep{yang22}). Whether a common origin exists for X, S, and Z-shaped RLAGN is currently a topic of discussion, and large sample statistics from heterogenous samples are needed to resolve this question.

On even larger scales, `giant' RLAGN, which are traditionally associated with sources exceeding $\sim$1\,Mpc in largest angular size, have often been used to select restarting sources (see, e.g., \citep{sari07} and references therein), as they are suggested to be at the tail end of the lifetime distribution and likely have a higher chance of showing restarted activity in the vicinity of the radio core. Those without radio cores are strong candidates for remnants as well, e.g., \citep{mach02}. There are suggestions that the large physical sizes of giants are themselves a result of multiple episodes of activity, where continued jet episodes expand the original remnant lobes further \citep{subr96,mach04}. Particularly prototypical examples of restarting sources selected from giant radio galaxy samples are J0116-473, a 2.1 Mpc source presented in \cite{sari02}, and PKS\,B1545-321 in \cite{safo08}. PKS\,B1545-321 has an inner double source size of 300\,kpc, while the outer remnant material is detected down to the radio core (at 1.5\,GHz;~\cite{safo08}), making it a prime example for use in studying the interaction between the restarting jet and the remnant lobe.

The causes of jet interruption and restarted activity are not yet understood. However, there is now robust evidence that restarted RLAGN are not hosted by a certain type of host galaxy relative to the RLAGN population \citep{maha19,jurl20}. Restarted activity is likely to be a normal part of the active radio lifecycle of massive ellipticals; thus, in the first instance, the central black hole and accretion properties that drive jet activity must be related to their modulation. There have been suggestions for quasars with the highest accretion rates that intermittency is favoured in elliptical galaxies (e.g., \citep{siko08,kuzm17}), or that stochastic development of poloidal magnetic fields in the accretion disks lead to intermittent jets~\citep{livi03,maye06}; nevertheless it is likely that episodic activity is intimately related to the activity of the accretion system in both quasars and non-quasars. Several mechanisms have been discussed to explain accretion disturbances, such as thermal instabilities \citep{prin76}, the infall of large gas clouds to the centre of the galaxy as a consequence of mergers~\citep{scho00}, and intermittent jet collimation by accretion disk winds \citep{siko07}; however, there is insufficient observational evidence for these processes occurring in restarting RLAGN as a population. While there are a growing number of cases of compact or young RLAGN with high molecular gas fractions (as discussed above), in \cite{sari07} a low CO detection rate was derived for a representative sample of restarting sources, suggesting instabilities in the fuelling/accretion process may be the cause of jet~disruption.

Understanding the timescales of the radio plasma between old and new activity can in a sense constrain the driver of jet interruption, as they can be compared to the timescales of modulation of accretion activity. Examples of DDRGs that contain hotspots in their outer lobes, explained by a case in which remnants of the old jet continue to arrive in the outer lobes, can constrain the timescale of interruption to a few Myr at most (e.g., B1834 + 620; \citep{scho00}). However, these cases are rare, and can only provide information on the jet duty cycle on a source-by-source basis rather than on the whole population. Despite their rarity, DDRGs have been widely used to determine jet duty cycles thanks to the clear distinction between episodes of activity. In this context, accurate modelling of the spectral energy distribution of the inner and outer doubles to obtain radiative age estimates requires sensitive maps across a broad frequency coverage. In \cite{jamr07}, the authors presented radio maps ranging from 240\,MHz to 8460\,MHz of the restarting source 4C\,29.30, deriving radiative age estimates of $\lesssim$33 Myr and $\gtrsim$200 Myr for the inner double and the outer extended emission, respectively. In \cite{mach10}, observations at frequencies between 334\,MHz and 4910\,MHz were used to study the DDRG J1548-3216 (PKS B1545-321), deriving radiative age estimates of $132\pm28$\,Myr and $\sim$$9\pm4$ Myr for the inner and outer doubles. Although the usual caveats about the fitting of spectral aging models to spectra (mentioned earlier in Section \ref{sect:remnants}) apply, these sources in particular have interruption times much larger than that implied by models of accretion variability (see the review by \citep{siem10} and references therein). A further study with similar spectral coverage of the DDRG J1453 + 3308 \citep{kona06} found a striking similarity between the inner and outer doubles for the low frequency injection index $\alpha_{inj}$, suggesting that the similarity arises in the case when the restarting and remnant jets are driven with the same jet power (confirmed by \citep{kona13} for multiple cases and modelling approaches). This might imply that DDRGs form when accretion is disturbed, but only under a scenario in which the accretion properties that drive jets are unchanged before and after the disturbance. This would seemingly be plausible and consistent with other studies if short-term instabilities temporarily halt jet production, rather than drastic changes in the external medium which may alter the magneto-ionic properties surrounding the accretion system thought to power magnetized jets.

\textls[-38]{Information on the true nature of restarted sources requires population studies; \mbox{Refs.~\cite{saik09,nand12,nand19}} have provided a collection of DDRGs that are not biased towards the largest Mpc scale sources, while \cite{kuzm17} presented a study of a sample of 74 sources with evidence of restarting activity compiled from the literature. This enabled an important investigation of empirical trends in observed radio and optical properties, as well as comparisons with a parent sample of (FR-II) sources. They found that, while black hole masses are similar between the samples, the stellar masses in restarting sources are lower. In addition, they found indications that restarting sources have evidence for merger history in their host galaxies over that found for FR-IIs. While these results remain important in the determination of the trigger for jet disruption, more robust information which confirms these trends can only be made with systematic selection (i.e., with flux and size completeness and sources selected in the same survey). As with the case of remnants, systematically selected samples of restarting RLAGN are scarce, mainly due to the difficulty in identifying restarting jets except in the case of DDRGs, although there are a number of candidate DDRGs that require follow-up observations for robust confirmation of their inner edge-brightened lobes. The advantages of using low frequency instruments to construct large samples of remnants apply to restarting sources as well, and such samples will allow future studies to infer statistical information on their properties. }

It should be borne in mind that restarted activity is not just restricted to RLAGN. Highly sensitive observations show that Seyfert galaxies, which rarely display luminous and/or large-scale jets, show nuclear radio flux variability in those galaxies where small (pc-scale) jets are not present \citep{mund11}; hence, there is a clear analogy with restarted RLAGN, for which core variability at a higher rate than normal RLAGN has been reported \citep{kona13_obs}. The authors of \cite{hota11} reported observations of a unique triple-double nature in MaxBCG J212.45357-03.04237, hosted by a spiral galaxy with recent star formation. In fact, there have been suggestions that the lack of radio-loud Seyfert galaxies may be due to a high fraction of intermittency in their jets rather than to low radio power \citep{fosc15}. This suggests that intermittency is a natural byproduct of jet production over a wide range of jet powers, and not just for the `tip of the iceberg' radio-loud sources that dominate flux-limited samples. Because lower frequencies sample the lower radio power domain, our knowledge of jet lifecycles is limited without further information on the whole AGN population.

\subsubsection{Low-Frequency Observations and Surveys}
\label{sect:lowfreq}
Low-frequency surveys are ideal for selecting remnant and restarting sources (a typical source is expected to be detectable for a few $\times10^8$ yr at 151\,MHz  \citep{cord86}). Hereafter, I refer to `low frequencies' as the regime of frequencies below 500\,MHz and above the atmospheric opacity of ground-based instruments detecting radio emission of $\sim$10\,MHz. Not only are low-frequency instruments sensitive to the steep-spectrum emission that is generally undetected by higher frequency instruments, in general they have the short baselines that are particularly important in observing the large-scale diffuse lobe material expected in remnant sources. Sub-arcminute angular resolution is generally required even for large angular size sources such as DDRGs; for a source at a redshift of $z=0.1$, arcminute angular resolutions can observe emission on scales of approximately 100\,kpc, which would provide ten beam-widths across a 1\,Mpc source. The caveat here is that low-frequency instruments do not generally have the long baselines required for detailed resolved studies on smaller sources; thus, low frequencies, particularly for statistical samples, are traditionally used for selection of steep-spectrum lobes and information on resolved structures is probed with follow-up observations at higher frequencies. Table \ref{tab:surveys} lists current low-frequency surveys up to 1\,GHz that are most suited for observing large samples of remnants and restarting sources at sub-arcminute resolution. I have included the VLA Low Frequency Sky Survey redux (VLSSr; \citep{lane14}), with a resolution of 75 arcsec but not far from arcminute resolution, as an important very-low-frequency survey.

\begin{table}[H]
\caption{Summary of low-frequency ($\leqslant$500 MHz) surveys at an angular resolution better than 1~arcmin (except for VLSSr), ideal for observing and selecting remnant and restarting populations. $^{\dagger}$~The LoDeSS survey is not yet complete, but the calibration techniques for future release have been presented by \cite{groe23}. \label{table:surveys}}
\newcolumntype{C}{>{\centering\arraybackslash}X}
\begin{tabularx}{\textwidth}{CCCC}
\toprule
\textbf{Name}	& \textbf{Frequency (MHz)}	& \textbf{Resolution ($''$)} & \textbf{RMS Noise \linebreak(mJy beam$^{-1}$)}\\
\midrule
LoDeSS $^{\dagger}$ \citep{groe23} & 23 & 45 & 12 \\
LoLSS \citep{dega23} & 54 & 15 & 1 \\
VLSSr \citep{lane14} & 74 & 75 & 100 \\
LoTSS \citep{shim22} & 144 & 6 & 0.083 \\
TGSS \citep{inte17} & 150 & 25 & 5 \\
WENSS \citep{reng97} & 327 & 54\,cosec\,$\delta$ & 3.6 \\
\bottomrule
\end{tabularx}
\label{tab:surveys}
\end{table}

\begin{figure}[H]
\includegraphics[scale=0.6]{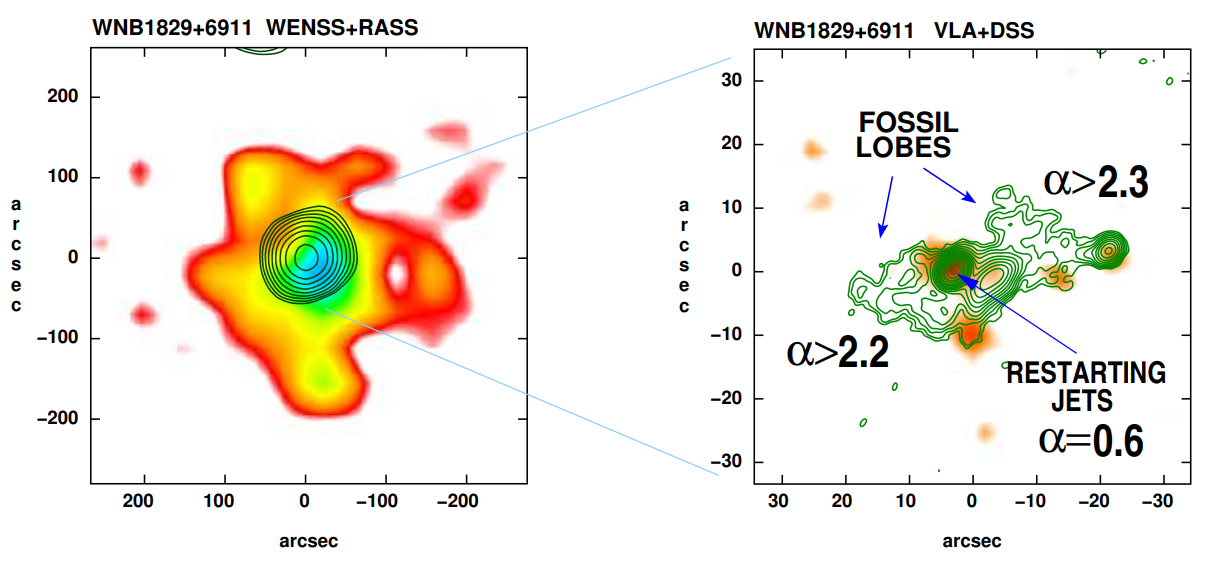}
\caption{The cluster-centre remnant source  of WNB1829 + 6911. \textbf{Left}: X-ray ROSAT All Sky Survey image with WENSS 327 MHz radio contours overlaid. \textbf{Right}: Digitized Sky Survey optical image with VLA 1.4 GHz radio contours overlaid, showing the outer steep-spectrum lobe material and the compact inner flat-spectrum source at the location of the host, with clear evidence of restarting activity. Image credit: \cite{murg05}. Reproduced with permission © ESO. }
\label{fig:murg+05}
\end{figure}

The Westerbork Northern Sky Survey (WENSS \citep{reng97}) minisurvey \citep{deru98} at 327\,MHz contained 3 of 400 sources with extremely steep low frequency spectral indices of $\alpha\geqslant1.7$, and drove follow-up high-frequency observations confirming their lack of jet activity on kpc scales \citep{murg05}. All three sources were found to be cluster-centre sources, providing more evidence of the scenario in which a dense medium prevents strong adiabatic losses and avoids the break frequency shifting to even lower frequencies. In their survey, Ref.~\cite{murg05} calculated the fraction of remnant sources in rich Abell clusters to be 60\%, whereas they found the fraction outside such environments to be 6\%. In other words, the remnant phase can last ten times longer in rich cluster environments, a conclusion similarly drawn \mbox{by \cite{murg11}}, confirming that while the spectral evolution of remnant lobes is extremely fast, it is slower in dense environments. An example of the type of steep-spectrum source that can be selected in such a survey is shown in Figure \ref{fig:murg+05}, possessing a flat spectrum core surrounded by steep-spectrum emission with amorphous morphology, and as such associated with restarting jets \cite{parm07,murg11}. Distorted and irregular radio morphology is common in these sources at low frequencies, even when the lobes are confined by high-pressure environments (e.g., NGC\,507; \citep{giac11}). Practically, it is difficult (and probably wrong) to assign remnant or restarting status to a source based solely on its low-frequency morphology. First, we do not have knowledge on its FR-I/FR-II morphological class when the source was active, and second, many sources with amorphous morphology have likely spent a large fraction of their lifetimes in the remnant state, making it difficult to assess their structural evolution (e.g., \citep{murg11}). I refer the reader to Section \ref{sect:summary} for a discussion on this topic. Similar methods utilizing survey data to select very steep spectral index sources and following up with higher-resolution data were used in a study by \cite{dwar09}, finding ten double-lobed sources without core detection, who then integrated spectral indices of $\alpha^{74}_{1400}\geqslant1.8$ using cross-matches between sources in the VLSS and NVSS surveys. The relative absence of genuine remnant and restarting sources in field environments feeds a requirement for even lower frequencies, higher resolution, and higher sensitivity than that allowed by the aforementioned surveys.

Recent developments in low-frequency instrumentation and calibration techniques have been key to furthering studies of remnants and restarting sources. Instruments such as the Low Frequency Array (LOFAR; \citep{vanh13}), the Murchinson Wide-Field Array (MWA; \citep{lons09}), and the Giant Metre Wave Telescope (GMRT) have been used extensively to image remnant and restarting sources to adequate resolution and sensitivity and in greater numbers. In~\cite{walk15}, a serendipitous discovery of the giant (700\,kpc) radio galaxy NGC\,1534 was presented using a 185-MHz image from the MWA at an angular resolution of $\sim$300 arcsec. The low angular resolution was required due to the extremely low surface brightness in the lobes of $\leqslant$$0.1$\,mJy\,arcmin$^{-2}$ at 1.4\,GHz. Such discoveries require deep follow-up at a broad range of frequencies and at adequate resolution in order to determine their synchrotron lifetimes and estimate their duty cycles. Indeed, \cite{duch19} used GHz-frequency follow-up observations to fit radiative models to the integrated spectra of NGC\,1534, estimating that the spectral age of the remnant emission is 203\,Myr, with the source only having been active for 44\,Myr. A similar study was performed on the NGC\,1407 rich galaxy group, hosting a restarting source active for 30\,Myr and the remnant emission for 300\,Myr, using observations between 240\,MHz and 1.4\,GHz with the GMRT and VLA~\citep{giac12}. In \cite{shul15}, LOFAR, GMRT, and VLA data (144--1425\,MHz) were combined to study the steep-spectrum source VLSS J1431.8 + 1331. By fitting resolved spectral ageing models to their data, a \textit{youngest} (equipartition magnetic field-based) particle population age of $\sim$60~Myr was determined, suggestive of fading but not necessarily ceased jets. A similar conclusion was made by \cite{rand20}, instead using a continuous injection model \citep{komi94}, for the source J1615 + 5452, finding a quiescent time of 21\,Myr, which was interpreted to be the time since particle acceleration in the lobes).

More robust age estimates require more accurate estimates of the lobe magnetic field strength, which are made possible with X-ray observations that describe inverse-Compton emission in the lobes. The main problem with this is that X-ray inverse-Compton detections are infrequent, particularly in very low radio luminosity sources. The best possibilities are in high redshift sources where CMB-driven inverse-Compton losses are stronger, leading to brighter emission, though this comes at the price of the radio emission being very weak. A rare example is the remnant J021659-044920 at a redshift of $z=1.3$, with radio lobes detected between 325\,MHz and 1.4\,GHz and lobe inverse-Compton emission detected in the X-ray band at 0.3--2.0\,keV \citep{tamh15}. In this particular case, the inverse-Compton and equipartition-based magnetic field estimates were consistent (which is not uncommon); moreover, the determination of the magnetic field allowed the determination of the total energy in relativistic particles of $4.2\times10^{59}$\,erg, a value  implying that even remnant lobes are capable of significant feedback into the surrounding medium.

Serendipitous discoveries are commonly made at low frequencies, with LOFAR at 150\,MHz as an example, made by \cite{brie16} for the source `blob1' (J2000.0 RA 18 h 28 min 20.4~s Dec + 49 d 14 m 43 s) with a derived time from `switch-off' of $t_{off}\approx60$\,Myr. In their study, blob1 was characterized as a remnant due to its very low surface brightness and amorphous morphology, though it contains a low-frequency spectral index consistent with typical active sources. Hence, selecting sources based only on their spectral index may not be reliable, and may miss potential remnants that have not necessarily switched off but have switched to a low-power state; e.g., while the radio core of blob1 has a spectral index of $\alpha^{4900}_{1400}=0.3$, its flux density is more than an order of magnitude lower than that expected for active galaxies~\citep{giov88}. Moreover, large-scale radio emission associated with remnant lobes may even become re-energized due to interactions with their environments (see the example of Nest200047 \cite{brie21}); thus, localized flat-spectrum emission may not necessarily indicate jet activity. The picture then arises that jets may not switch off completely, instead switching to a low power on Myr timescales such that the resulting radio emission dims beneath detection limits.

\begin{figure}[H]
\newcolumntype{C}{>{\centering\arraybackslash}X}
\begin{tabularx}{\textwidth}{L}
\includegraphics[scale=0.5]{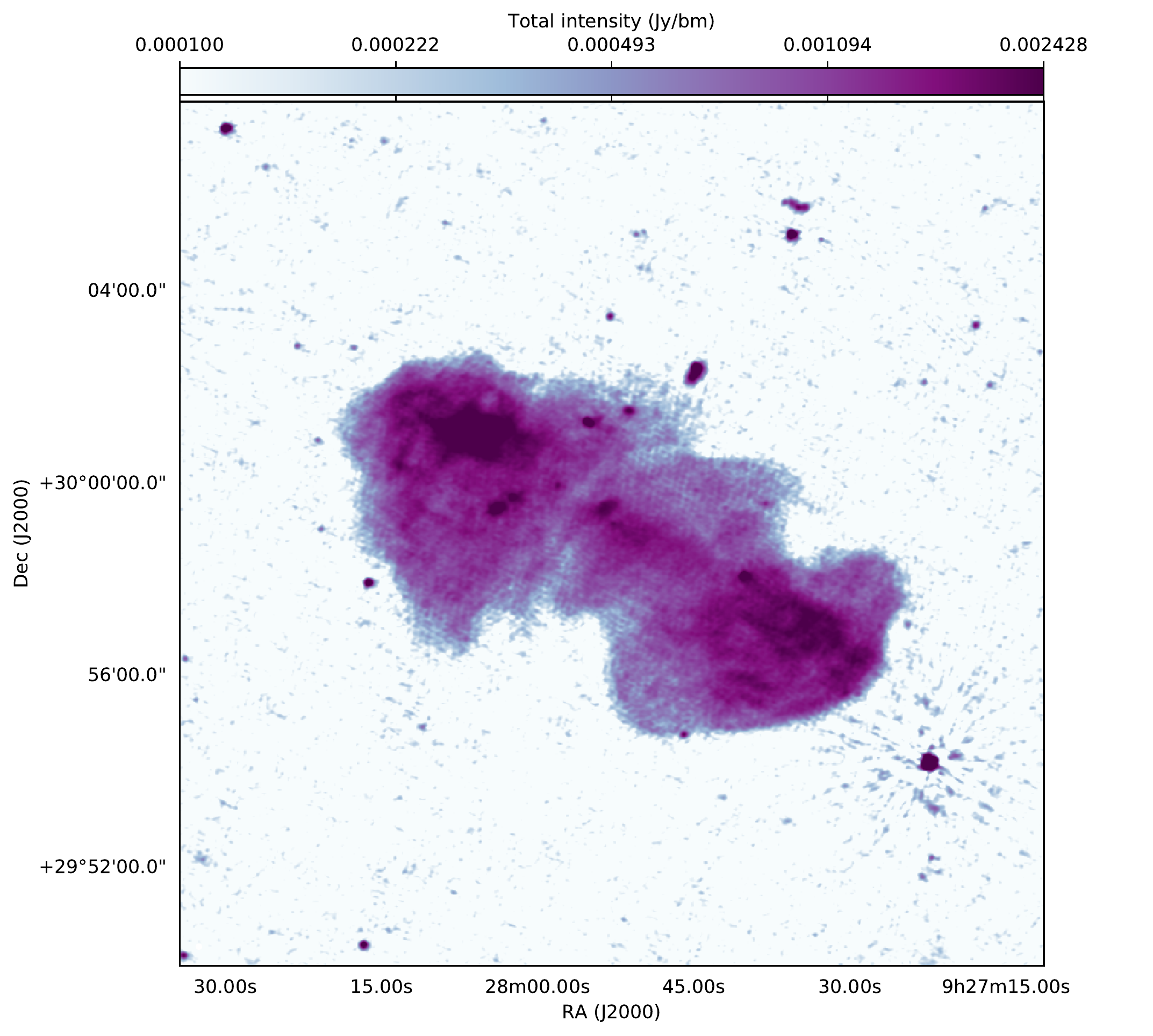}\\
\includegraphics[scale=0.5]{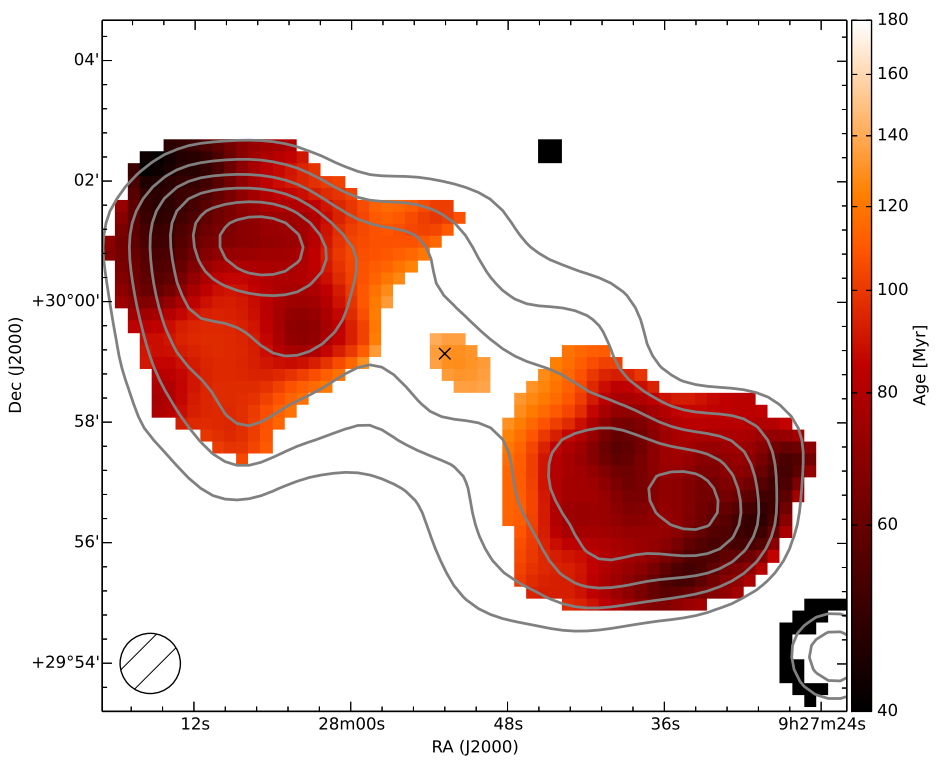}\\
\end{tabularx}
\caption{The well-studied remnant source B2\,0924 + 30. \textbf{Top}: 144\,MHz total intensity map observed in the LoTSS survey at 6 arcsec resolution, from the publicly available archive of DR2, accessed in January 2023.  
\url{https://lofar-surveys.org/dr2_release.html}, accessed on 25 May 2023. \textbf{Bottom}: radiative age map by \cite{shul17} at 1~arcmin angular resolution.}
\label{fig:shul+17}
\end{figure}

In \cite{shul17}, the authors studied the well known remnant source B2\,0924 + 30, using frequency coverage down to 113\,MHz at sub-arcminute resolution, providing the spectral \textls[-15]{age map shown in Figure \ref{fig:shul+17}. As seen in the figure, and as stated by the earliest identification~\citep{cord87}}, the source lacks compact emission on scales down to 1 arcsec. The regions of younger age towards the edges of the source ($\gtrsim$50\,Myr), which are suggested to be remnants of the hotspots, class the source as a fading FR-II. Such broad-band spectral studies have helped to determine radiative lifetimes of the inner and outer doubles of DDRGs, as presented \mbox{by \cite{kona12}}, finding that the remnant phase can be as short as a few per cent of the active phase in the DDRGs they study. In \cite{orru15}, early LOFAR commissioning data on the \mbox{DDRG B1864 + 620} (between 144\,MHz and 8460\,MHz) were used, finding that both the northern inner and outer doubles remain active, suggestive of a very short timescale between interruption of activity. A similar conclusion was made through LOFAR observations of the source B2\,0258+35, suggested as being a restarted source based on its morphology \citep{shul12}; however, broad-band spectral information between 145\,MHz and 6600\,MHz suggested that the lobes continue to be being fed by a jet \citep{brie18}. In \cite{brie20}, 144\,MHz observations of the famous restarting source 3C\,388 were used, relying on single-injection radiative models to find a duty cycle of $\frac{t_{on}}{t_{on}+t_{off}}\gtrsim60\%$ during a total synchrotron lifetime of $\lesssim$80\,Myr, which is a duty cycle far larger (or an interruption time far shorter) than that found for DDRGs with similar spectral coverage.

Information on the population of remnants and restarting sources are needed to answer the most fundamental questions, such as whether remnants truly have a radio jet or switch to a low-power state, whether remnant lobes have the energetics to contribute to AGN feedback, and what the causes are for interrupted and restarting activity. A more stream-lined question is: what determines radio-loudness? Answering this question requires statistical information on systematically-selected samples, rather than detailed studies on single sources. Very few instruments have the capabilities, particularly in survey mode, to observe and detect sources at low frequencies ($\sim$100\,MHz), with very low surface brightness, at high angular resolution (tens of beam-widths across the source), and with $uv$--coverage adequate to observe the large-scale and diffuse emission expected of remnant lobes.

\begin{figure}[H]
\includegraphics[scale=0.37, trim={1cm 0 0.5cm 0},clip]{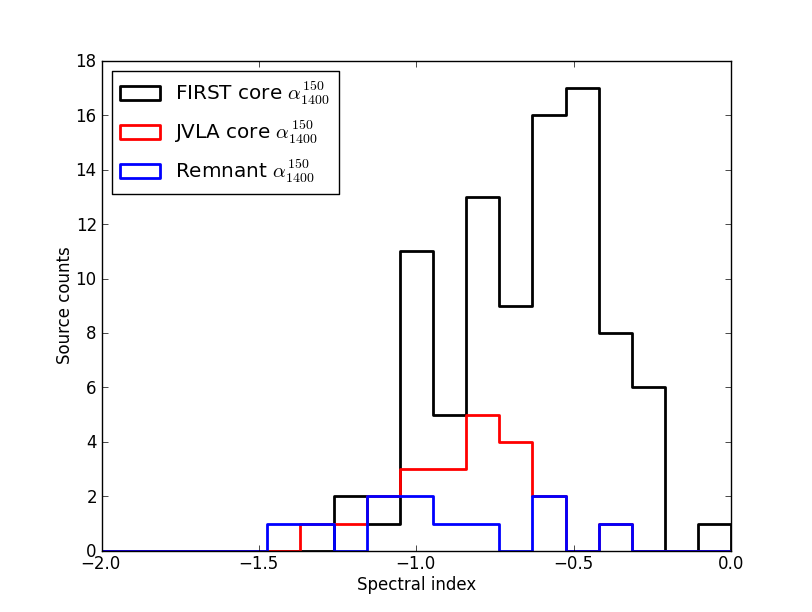}
\includegraphics[scale=0.2]{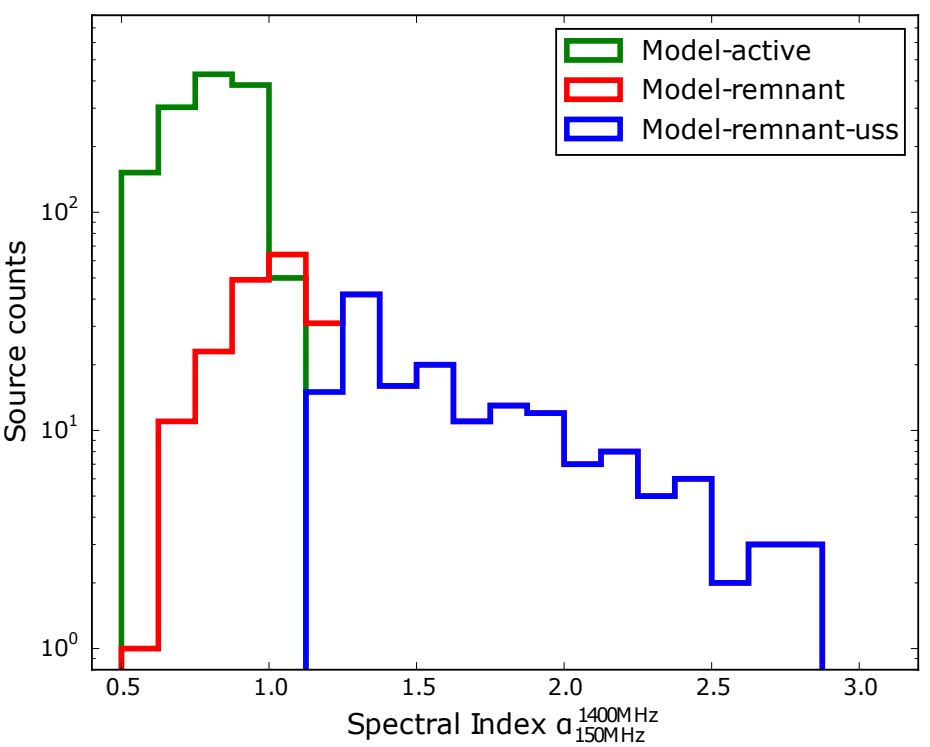}
\caption{\textbf{Left panel}: distribution of lobe spectral indices ($S\propto\nu^{\alpha}$ here, not $S\propto\nu^{-\alpha}$ as for the right panel) between 150\,MHz (LOFAR) and 1400\,MHz (NVSS) for the candidate remnant sample selected by \cite{hard_lofar16}. Blue: candidate remnants with no high frequency core \citep{maha18}. Red: candidate remnants with a high frequency core that was not detected by FIRST. Black: initial flux-complete sample with FIRST cores (see text). \textbf{Right panel}: distribution of lobe spectral indices between 150\,MHz and 1400\,MHz for the mock radio galaxy catalogue using radiative and dynamical evolution for active (green), remnant (red), and remnant with ultra-steep spectra (blue) from \cite{brie17}. }
\label{fig:remnant_dists}
\end{figure}

\textls[-18]{The most genuine remnants, with a combination of features suggestive of a lack of AGN activity (i.e., lack of compact structure, steep-spectrum lobes and relaxed morphology; see the extremely rare example of galaxy-scale remnant lobes in J122037.67 + 473857.6; \citep{webs21}), seem to be even rarer than DDRGs, even though DDRGs or restarted activity in general should be the more rare phenomenon. LOFAR, which has the ability to produce images at an angular resolution between 0.2 arcsec and $\sim$few arcmin or at frequencies between 10 MHz and 200~MHz, has been used extensively for statistical and complete studies of low surface brightness and steep-spectrum objects over wide sky areas. In \cite{hard_lofar16}, LOFAR observations were presented for the Herschel-ATLAS field covering 142 deg$^2$ at 150 MHz at 6 arcsec resolution, detecting 15,292 sources. The authors selected 127 RLAGN from the survey with a flux density $\geqslant$80~mJy and an angular size 40 arcsec, with associated optical hosts classed as AGN on the radio-far-infrared correlation. Of these objects, 38 had no visible evidence for a radio core, representing a candidate remnant fraction of 30\%, clearly higher than core-less objects in 3CRR. Higher frequency (6\,GHz) and higher resolution (0.3 arcsec) follow-up observations of this sample of 38 constrained the candidate remnant fraction in the H-ATLAS field to 9\%~\citep{maha18}, comparable to previous studies, further suggesting that the remnant phase is indeed rapid and can escape detection in even the most sensitive low-frequency surveys. Their sample consisted of a broad spectral index distribution ($0.4\leqslant\alpha_{144}^{1400}\leqslant1.5$; see left panel of Figure~\ref{fig:remnant_dists}), suggesting that although steep-spectral indices may select `genuine' remnants, recently switched-off sources that are just beginning to fade are missed by such selection criteria. Of course, these candidate remnants may indeed display a radio core with even more sensitive observations, but the question remains whether this faint core or core-less population represents a subpopulation of low-jet power sources or genuine remnants.}

A similar study performed by \cite{quic21} used a large frequency coverage between 119\,MHz and 9500\,MHz to select core-less RLAGN in the Galaxy and Mass Assembly (GAMA~\citep{driv11}) 23 field, finding ten candidates, most of which display hotspots, but only one with ultra-steep spectra, representing a remnant fraction of $4\%\leqslant f_{rem}\leqslant10\%$. Their broad-band coverage allowed robust spectral modelling, showing that hotspots can persist for up to 10\,Myr (in their sample) after jets switch off, meaning that remnants that still have hotspots may represent a significant fraction of genuine remnants. If a radio survey is sensitive to remnant plasma with an upper-limit age of $\leqslant$$10^8$\,yr (plausible at low frequencies), then their study indicates that at most 10\% of genuine remnants may have hotspots. Nevertheless, remnant fractions based on samples selected due to the lack of a radio core are only upper limits, as radio cores could clearly be detected at higher sensitivity than that achievable by any particular instrument. The future SKA, which may reach nJy sensitivity levels at GHz frequencies (discussed in Section \ref{sect:nextgen}), will shed further light on this; the key solution to this problem is to determine the limiting radio luminosity at which radio emission from the core is expected to describe a pc-scale jet driven by an AGN.

\textls[-15]{Using LOFAR observations of the Lockman-Hole field, Refs. \citep{maho16,brie17} used a combination} of spectral and morphological criteria, as the application of only one criteria clearly has biases \citep{godf17}. They selected only those sources with an ultra-steep spectral index ($\alpha^{1400}_{150}\geqslant1.2$) and high spectral curvature (SPC $\alpha^{1400}_{325}-\alpha^{325}_{150}$, where SPC$\gtrsim0$ can select sources with spectral indices that are flat at low frequencies and have significant steepening at higher frequencies) or those sources with an absence of a core or hotspots and with relaxed morphology, identifying 23 candidate remnants (later refined to 13 by \cite{jurl21} for a remnant fraction of 7\%). Although there is incompleteness triggered by independent sampling with different criteria, the remnant fractions from each selection process ranged from 4.1\% to 30\%. However, the benefit of these selection processes is the likelihood of selecting candidate remnants with a variety of properties related to different stages of remnant plasma evolution; the ultra-steep criterion is suggested for selection of the oldest remnants, while the lack of a radio core (though with bright lobes and even hotspots) can be used to select the youngest remnants. In \cite{brie17}, mock catalogues of low-power RLAGN were simulated using radiative and dynamical models to predict the number of remnants expected to be detected in the Lockman-Hole field, finding that the simulated remnant fraction of ultra-steep spectrum sources ($\alpha^{1400}_{150}\geqslant1.2$) was 10\% (see right panel of Figure~\ref{fig:remnant_dists}) which, including prescriptions for the lobe dynamical evolution, is consistent with the authors' upper-limit observed remnant fraction for ultra-steep spectrum low-power sources of $\leq$$15\%$. When modelling only for the radiative losses (i.e., without adiabatic losses in the lobes), the simulated remnant fraction was 24\%, significantly higher than actually observed, highlighting the importance of dynamic evolution in aged remnant lobes. It is clear, however, as shown in the right panel of Figure \ref{fig:remnant_dists} (from \cite{brie17}), that remnants which have ultra-steep spectra do not dominate remnant samples even at frequencies as low as 150\,MHz observed with high sensitivity. The models predict that remnants will scarcely be observed at redshifts greater than $z=1$ due to stronger inverse-Compton losses, meaning that cosmological evolution studies of remnants or restarting sources will not be possible even in the near future. Owing to sensitivity issues, particularly at higher frequencies, radiative models employing new techniques to analyse observed spectra for robust age estimates continue to be developed, and hold promise \citep{turn18,quic22}.

In \cite{jurl20}, previous studies of the Lockman-Hole field were complemented with more sensitive LOFAR data (RMS noise of 28\,$\upmu$Jy\,beam$^{-1}$, compared to $\sim$160 $\upmu$Jy\,beam$^{-1}$ in the initial study \citep{maho16}) to further select candidate restarting sources. The authors selected sources with a core prominence (CP; ratio of the luminosities of the core and of the lobe) of $\geqslant$$0.1$ and steep-spectral index of the radio core ($\alpha^{150}_{1400}\geqslant0.7$, as this is suggestive of newborn jets;~\citep{odea98}) or sources which are obvious DDRGs. Their restarting fraction of 15\%, generally larger than remnant fractions found in previous studies, suggests that jets restart quickly after the remnant phase in these objects.

With the resolution and sensitivity of NVSS as limiting factors in detection and spectral index statistics in the previous studies, Ref. \cite{morg21} combined LOFAR images of the Lockman-Hole field with wide-field images from Apertif \citep{oost18,adam19} at 1400\,MHz (15 arcsec resolution and RMS noise of 30 $\upmu$Jy\,beam$^{-1}$, significantly better than NVSS) to produce the first wide-field spectral index map at high resolution. They determined remnant and restarting fractions of 9\% and 7\%, respectively, consistent with predictions from simulations of mock catalogues~\citep{brie17}. Thus, low frequency-selected samples of remnant and restarting sources represent at most 10\% of the RLAGN population. The insights presented by the aforementioned studies are crucial for current and future wide-area low-frequency surveys.

\begin{figure}[H]
\newcolumntype{C}{>{\centering\arraybackslash}X}
\begin{tabularx}{\textwidth}{L}
\includegraphics[scale=0.4]{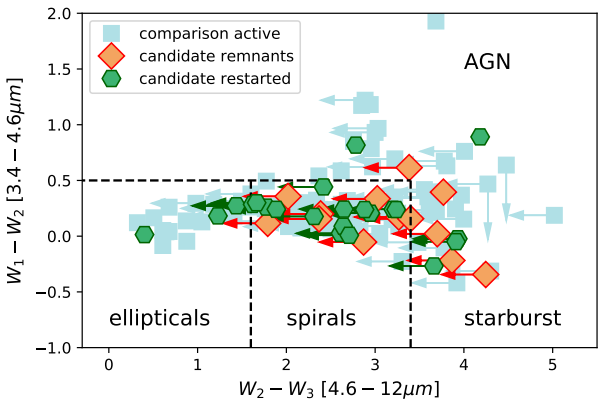}\\
\includegraphics[scale=0.3]{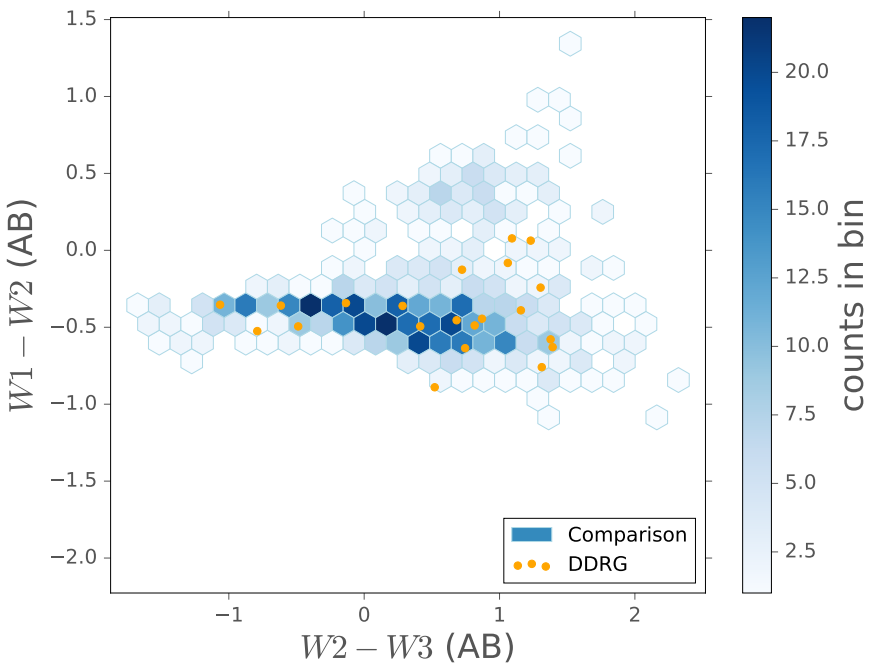}\\
\includegraphics[scale=0.4]{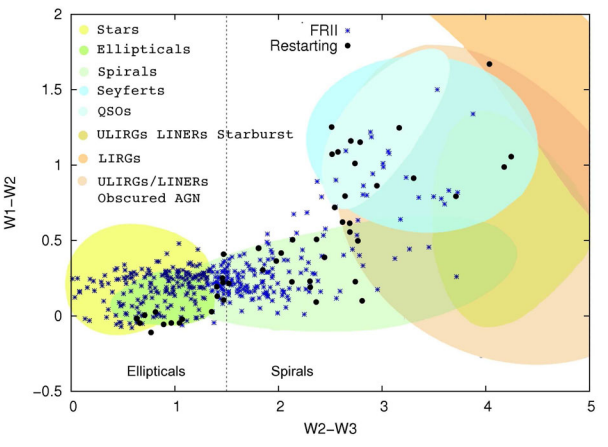}\\
\end{tabularx}
\caption{WISE colour--colour plots for the host galaxies of remnants and restarting sources (\textbf{top}~\cite{jurl20}; \textbf{bottom} \cite{kuzm17}) and DDRGs (\textbf{middle} \cite{maha19}) compared with those for the parent samples of RLAGN. Both studies that select samples heterogeneously from the same survey but with different selection methods (top and middle) replicate the conclusion that interrupted radio jet activity is not a consequence of being hosted by different types of galaxies, a conclusion based on optical and infrared photometric~properties.}
\label{fig:colourcolour}
\end{figure}

\textls[-18]{The aforementioned studies represent useful tests to assess the efficiency of selection strategies for use in large surveys, such as the LOFAR Two-metre Sky Survey \linebreak (LoTSS;~\citep{shim19,shim22}) of the entire northern sky. The combination of wide sky area (all 2$\pi$ steradians of the northern sky), angular resolution (6 arcsec), and sensitivity (median of $\sim$83~$\upmu$\,Jy\,beam$^{-1}$) makes large and complete statistical samples of remnant and restarting sources possible, as well as serendipitous discoveries (e.g., \citep{savi18,brug18}) undetected by past surveys. Using Data Release 1 (DR1; \citep{shim19,will19,dunc19}), covering 424 deg$^{-2}$ of the HETDEX field, Refs. \citep{hetdex,maha19} constructed a sample of 33 DDRGs, robustly identified using higher resolution follow-up observations to confirm that the inner doubles were edge-brightened. By comparison, with a control sample of `normal' (non-restarting) RLAGN from the same survey in \cite{hard19}, they found that the host galaxy photometric properties (e.g., near infrared colours, optical magnitudes) were statistically indistinguishable between DDRGs and the radio galaxy population, suggesting that restarted activity is not driven by different host galaxy types. This result was further confirmed by \citep{jurl20} for their selection of restarting sources of various morphology (i.e., not just DDRGs) as well as for remnant sources (see Figure \ref{fig:colourcolour}). The only explanation for the driver of restarted activity is constrained to the physics on scales smaller than those that are probed by photometric measurements by current optical and infrared surveys.}

\begin{figure}[H]
\includegraphics[scale=0.4]{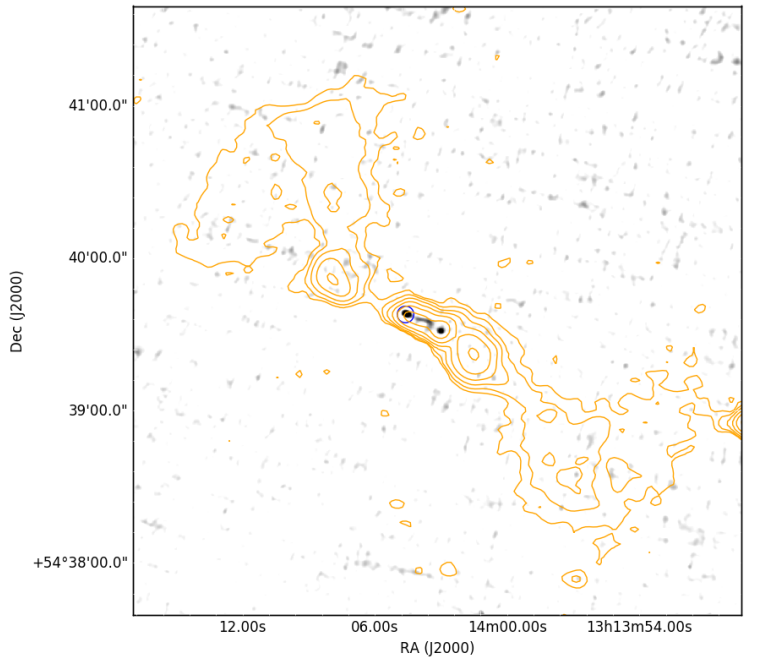}
\caption{The source ILTJ131403.17 + 543939.6 from the DDRG sample of \cite{maha19}, showing evidence of a third episode of activity based on the diffuse extended plasma beyond the outer double. Greyscale image: 6\,GHz VLA data showing the inner restarting double. Orange contours: 144\,MHz LOFAR~data. }
\label{fig:tripledouble}
\end{figure}

Highly resolved spectroscopy that may provide measurements or estimates of accretion rates or information on emission lines in these galaxies in general may provide further insights into the differences of host galaxies between restarted, remnant, and active sources. The upcoming WEAVE-LOFAR survey \cite{smit16} should shed further light on this longstanding unknown, although SDSS spectral analysis between remnant, restarting, and active RLAGN performed by \cite{jurl21_spectral} found no difference in the optical emission line properties of their hosts, suggestive of similar accretion levels. Meanwhile, in light of the short baselines attributable to low frequency surveys, a number of candidate \textit{triple} double RLAGN showing three episodes of activity have been found: an inner double, outer double, and diffuse amorphous plasma connected with and extending beyond the outer double (Figure \ref{fig:tripledouble}). The existence of these objects provides unambiguous evidence that restarting activity is not a one-time occurrence in RLAGN, and that many (or all) double-lobed RLAGN may themselves be restarted activity. Considering the small number of probable triple-double sources (see further examples by \cite{broc07,mare23} and the review by \cite{saik09}), and that DDRGs themselves are rare, selection of such sources for study requires even lower frequencies (e.g., LoLSS \cite{dega21}).

Other than increasing number counts, there have been other advantages to sensitive low-frequency surveying. The large populations of remnant and restarting sources has allowed the calibration of dynamical models to determine duty cycles (\citep{shab20}, discussed further in Section \ref{sect:simulations}). Another utility of broad-band surveying or imaging to low frequencies is understanding the radio morphology of sources selected at frequencies above the radio domain; for instance, hard X-ray-selected giant radio sources with Mpc-scale lobes have been shown to have a high fraction of self-absorbed cores (e.g., GPS sources), indicative of restarted activity \citep{brun20}. This has led to the idea that giant radio sources, due to their expected longer dynamical lifetimes, are more likely to be remnant or restarting sources, although deep low-frequency imaging reveals structures such as wings (which do not necessarily mean restarting) in these objects \citep{brun21}.

The main conclusions of recent statistically-complete studies of remnants and restarting sources at low frequencies can be summarized as follows:
\begin{itemize}
\item Robust remnant and restarting source fractions are $\sim$10\% at most;
\item Restarted activity is a very rapid phenomenon;
\item Remnants with radiative ages as old as a few hundreds of Myr have been found, particularly those sources with very steep spectral indices;
\item Certain genuine remnants may not have very steep spectral index lobes, such as recently switched-off sources that may have hotspots, as the radio structure has strong evolution in the remnant or restarting phase;
\item Disrupted jets are not associated with a particular type of host galaxy.
\end{itemize}

\begin{figure}[H]
\includegraphics[scale=0.5, trim={0 0 0 5cm},clip]{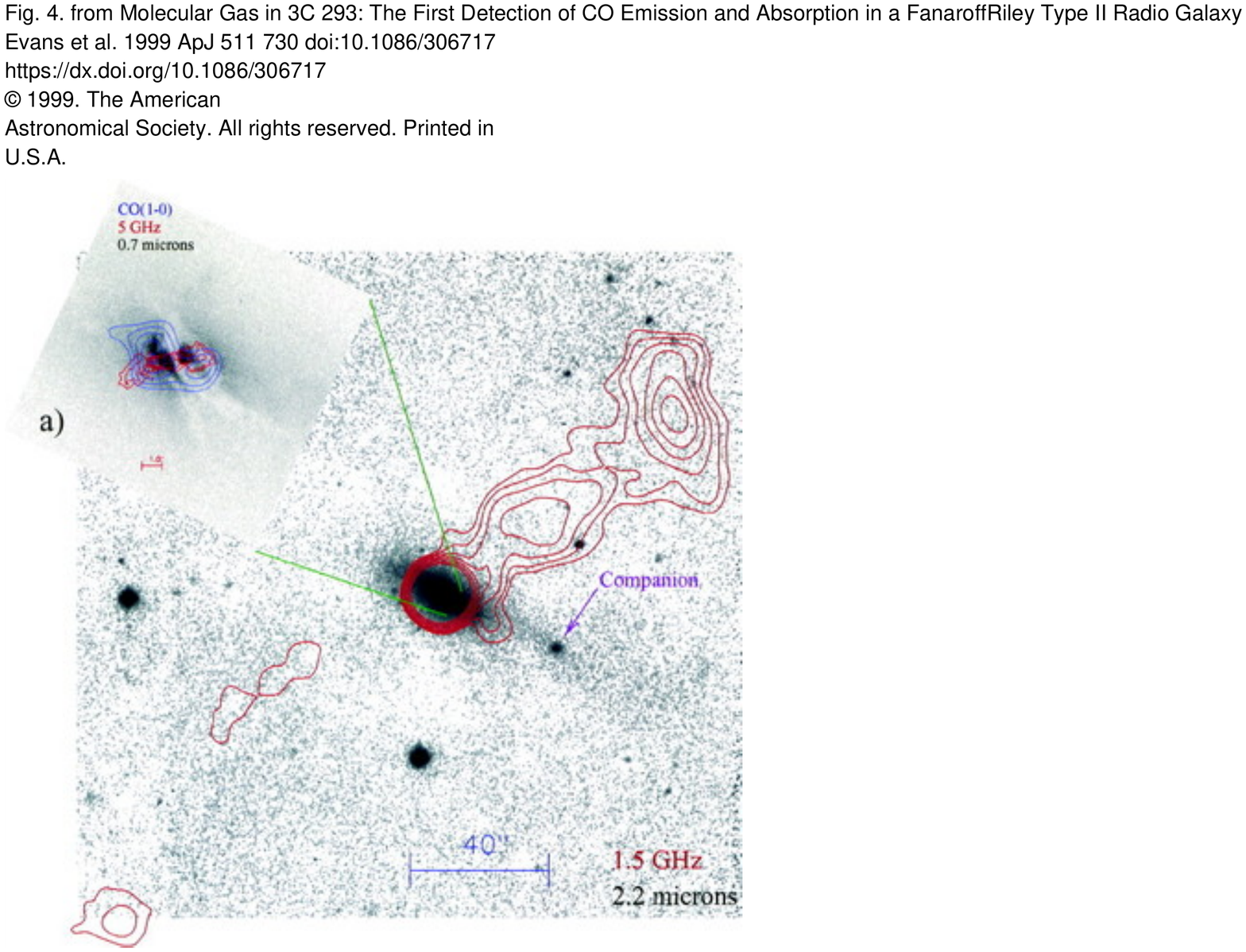}
\includegraphics[scale=0.9]{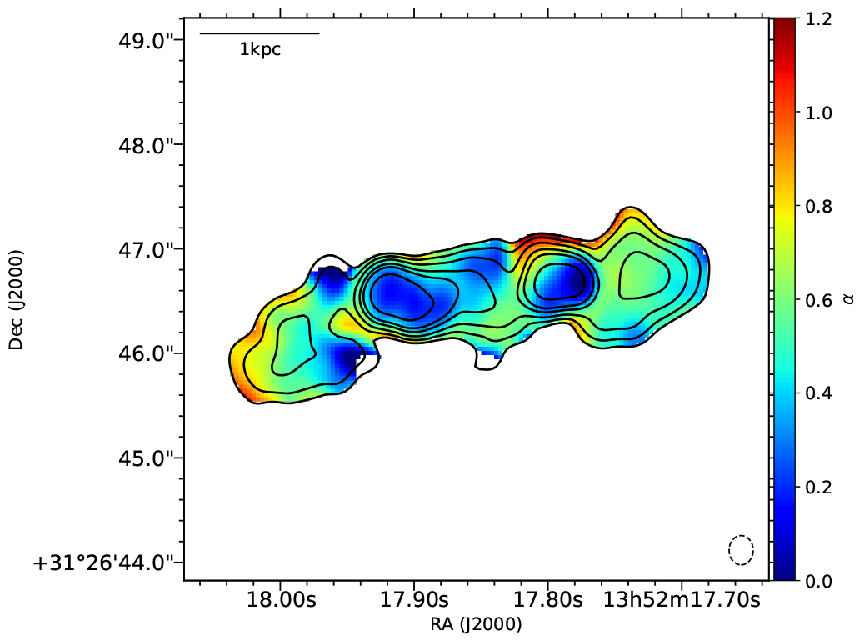}
\caption{\textbf{Top}: 
1.5\,GHz radio image (red contours) of 3C\,293 superposed on a near-infrared 2\,$\upmu$m image (from \cite{evan99}). The inset (a) shows 5\,GHz MERLIN contours (red), with the inner restarting jet and CO distribution showing the molecular gas (blue) overlaid on a \textit{HST} image with large dust lanes (© AAS. Reproduced with permission). \textbf{Bottom}: Spectral index map between 144\,MHz and 1360\,MHz at 0.28 $\times$ 0.23 arcsec resolution (from \cite{kukr22}), clearly showing the distinct diffuse older electron population beyond the flat spectrum inner lobes. }
\label{fig:3c293}
\end{figure}

There are few sources that show a combination of all the suggested properties of restarting sources. One of the most well-studied sources, 3C\,236, is known to have elongated edge-brightening in the northern lobe \citep{bart85,shul19}, repeated bursts of star-formation in its host galaxy \citep{trem10}, an inner CSS source \citep{trem10}, and fast outflows of cold molecular gas and atomic hydrogen \citep{morg05,schu18}, all of which are associated with young radio jets. Another example is 3C\,293, which is known to have asymmetric large scale lobes at a size of $\sim$220 kpc and a bright steep-spectrum core on arcsec scales, though on sub-arcsec scales the core is resolved into a double-lobe structure $\sim$4.5 kpc in size \citep{brid81} on either side of a flat-spectrum core~\citep{akuj96}. With the source being hosted by a post-merger galaxy with multiple dust lanes~\citep{cape00} and large amounts of cold gas in the region of the inner double (incidentally, the first detection of CO emission in an FR-II source was in 3C\,293 by \cite{evan99}; see top panel of Figure \ref{fig:3c293}), 3C\,293~\citep{evan99,besw02,ogle10} is another prototypical example with multi-wavelength evidence of restarted activity. Recently, further strides have been made towards harnessing the full international array of LOFAR, providing baselines of $\gtrsim$$2\times10^6\lambda$. Calibration of the international stations with the High Band Antennas (HBA) at a central frequency of 144\,MHz \citep{mora22} has facilitated further spectral ageing studies at an unprecedented angular resolution down to $\sim$$0.2$ arcsec. In \cite{kukr22}, such observations were used to provide highly resolved spectral aging maps of 3C\,293 down to 0.2 arcsec scales (see the spectral index map in the bottom panel of Figure~\ref{fig:3c293}), confirming the restarted nature of the inner double with a spectral age of $\lesssim$$0.17$\,Myr, which is a CSS source that is absorbed due to a rich ISM.

There has been tremendous progress in recent years when it comes to obtaining insights into RLAGN lifecycles; however, our understanding remains limited by the most fundamental question of how restarted activity is triggered. Possible questions for the future include:
\begin{itemize}
\item Is episodic activity related to instabilities or episodic accretion flow, and how sensitive is the consequent jet production to changes in the accretion system?
\item Are remnant lobes capable of AGN feedback, and do they contribute to the existence of diffuse cluster sources?
\item To what extent does the large-scale environment play a role in episodic behaviour?
\item Has every RLAGN passed through the remnant and restarted phases?
\end{itemize}

We cannot hope to solve these questions with radio observations alone, and a complete picture of all remnant and restarting systems requires knowledge of other radiation mechanisms and their environments obtained by probing of multi-wavelength data.

\subsection{Multi-Wavelength Observations}
\label{sect:multiwavelength}
RLAGN predominantly emit synchrotron and inverse-Compton emission while driving shocks in the ambient hot gas medium that can be detected at frequencies higher than in the radio regime. This has facilitated many studies over the years attempting to understand the energetics of the hotspots, lobe magnetic fields, bow shocks in the surrounding medium, and interaction between the radio lobes and their environments (see review by \citep{hard20} and references therein). What are the multi-wavelength properties of remnant and restarting RLAGN, and what can they tell us about the role of RLAGN life-cycles in galaxy evolution?
\subsubsection{X-rays and Gamma Rays}
\begin{figure}[H]
\includegraphics[scale=0.5, trim={0 0cm 0 0cm},clip]{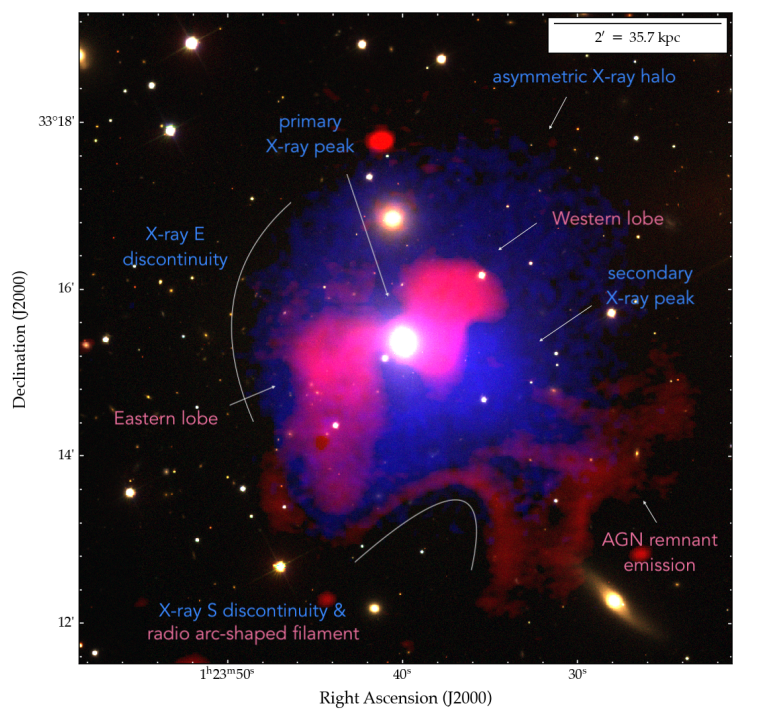}
\caption{Composite optical (background), X-ray (blue), and radio (red) image of the galaxy group NGC\,507 (from \cite{brie22}). Remnant emission newly detected by LOFAR, interpreted as remnant emission from a previous AGN episode, is suggested to be transported by gas sloshing induced by merger events. Reproduced with permission © ESO. }
\label{fig:brienza+22}
\end{figure}

\textls[-25]{X-rays provide important information on the large-scale (thermally-emitting) environment} that can provide clues toward the interaction between lobe plasma and its immediate surroundings. The prototypical remnant source NGC\,507 (or B2\,0120 + 33; \mbox{see \citep{giac11,murg11}} \textls[-15]{mentioned earlier in Section \ref{sect:lowfreq}) was further studied by \cite{brie22}, combining deep low-frequency} radio data and X-ray data describing the large-scale environment. This spectacular example of remnant radio emission represents one of the clearest examples of the interaction between remnant lobes and the complex X-ray emitting gas in the ICM (see Figure \ref{fig:brienza+22}). The newly detected remnant emission (arc-shaped filament and diffuse emission to the southeast of the central AGN) unambiguously show the remnant plasma being redistributed by the disturbed ICM in this system. In particular, referencing to Figure \ref{fig:brienza+22}, remnant emission from the eastern lobe is seen to `leak' and follow the X-ray discontinuity (interpreted as cold fronts in the ICM), likely being displaced by gas sloshing driven by merger events. Similar processes have been inferred in Abell\,3560 \citep{vent13}, highlighting the intimacy between diffuse remnant plasma and cluster processes. It is clear that massive clusters have the means to distribute low-density remnant plasma from RLAGN. In addition, there is a tendency for the cluster environments of remnant and restarting sources to be `cool-core' and to have high metal abundance (e.g., \citep{slee01,murg11}), potentially reflecting the AGN feedback cycle. For our close neighbour M\,87, repeated outbursts as long as every $3\times10^7$\,yr \citep{jone04} are required to quench the cooling flow of its gaseous atmosphere from shock heating alone.

Restarted jets are believed to be propagating not into the surrounding IGM/ICM but through the remnant lobes. As a result, models predict that moderate shocks are driven by the impact between the jets and the remnant plasma (see discussion in Section \ref{sect:simulations}). There are systems, although very few, where restarted jets are driving into the external environment and are detached from the outer remnant plasma while shock-heating the ISM/IGM/ICM (4C\,32.36; \cite{jeth08}, Cen\,A; \cite{kraf03,kraf07}). Such systems show evidence for multiple shocks in the hot gas medium from multiple episodes of jet activity; a prototypical example is from the multiple cavities and ripples seen in Perseus A/3C\,84 \citep{gend20}. These systems offer the best cases to study AGN feedback in different stages of their lifecycle.

In terms of the jets, lobes, and hotspots, X-ray emission is non-thermal and is generally attributed to inverse-Compton (IC) scattering (e.g., \citep{felt69,cros05}), for which the background photon field can be the CMB, synchrotron photons from the relativistic plasma itself, galaxy stars, or from the central AGN. On large scales, remnant and restarting sources are not expected to contain hotspots and jets; thus, the lobes are expected to radiate IC emission at some level from the CMB (e.g., as detected in the lobes of certain DDRGs \citep{kona19}). As we now know owing to numerous studies (detailed above in Section \ref{sect:lowfreq}), the radio emission at GHz frequencies is rapidly fading and usually undetectable for the remnant plasma; however, this is advantageous for IC X-rays. For soft X-rays ($\sim$1\,keV), electrons with energies of $\gamma\sim10^3$ are required to radiate through the IC process. As these are relatively low-energy electrons, radiating at $\lesssim$100\,MHz, their loss timescales are far longer than higher energy electrons (see the case of the restarting source 4C\,23.56, with X-ray IC emission pointing to remnant activity \cite{blun11}, or the strikingly similar 4C\,29.30 \cite{siem12}). The term IC `ghosts' (e.g., \citep{sara98,birz04,clar05,mocz11}) reflects this state in which there are IC X-ray detections of plasma that are not associated with current radio-emitting plasma from AGN unless the same $\gamma\sim10^3$ electrons are observed at the lowest radio frequencies. There are such cases of X-ray cavities in clusters that are not spatially associated with observed radio lobes, and which are attributed to buoyantly rising (undetected) remnant lobes, such as in the cases of Abell\,2597 \citep{mcna01} and NGC\,1275 (Perseus A/3C\,84) \citep{fabi00}. This physical process allows constraints to be placed upon the timescales of activity, as cavities without internal pressure support would collapse on the sound-crossing timescale (e.g., $\geqslant$$10^7$\,yr for Abell\,2597; \citep{mcna01}). Moreover, as the CMB energy density scales to $(1+z)^4$, the dimming due to distance is effectively cancelled out, and the remnant plasma is detectable with soft X-rays out to high redshift (more remnants are predicted to be detected this way by a factor of ten compared to radio observations at $z>2.2$, and by a factor of a hundred compared to $z>3.1$ \citep{turn20_xray}). Even if X-ray IC cannot be detected, upper limits are important to provide constraints on magnetic field strengths and the particle content of remnant plasma (e.g., for the remnant 0917+75; \citep{harr95}). On the other hand, ghost cavities, although suggested to contain enough energy to counteract cooling, may not originate from the lobes of remnant RLAGN on energetic grounds (NGC\,741; \citep{jeth08_cavities}), which otherwise require extreme physical parameters. It is possible that in these systems the remnant plasma evolves significantly in terms of dynamical evolution and particle content as injection, such that their energetics may not be comparable to our current understanding of the properties of radio lobes.

For hard X-rays (above 5\,keV), emission usually originates in the core, likely tracing components close to the AGN system, such as the hot corona. While hard X-ray selected samples of RLAGN are only biased toward sources with radiatively efficient accretion, strong nuclear X-ray luminosities can point towards restarted activity as compared to radio luminosities of any extended lobe emission, which has particularly been found to be the case for giant radio sources~\citep{ursi18}. Similar trends are seen in the gamma-ray band \citep{ursi19}. For gamma rays, inverse-Compton emission in the TeV band is predicted to be detectable for shocked shells surrounding the remnant lobes of 3C\,84 \citep{kino16}. If such a detection comes into fruition, this will enable study of high energy physics in remnants for the first time.

\subsubsection{Optical and Infrared}
While optical and infrared observations are useful for studying the particle acceleration in the jets and hotspots of the most luminous sources, for remnant and restarting sources the most important use is to have a host identification. However, this is particularly difficult for remnants that lack a radio core. Baldwin, Phillips, and Terlevich (BPT) diagrams are a diagnostic tool to identify AGN activity, and can be useful to determine signs of an AGN through the presence of narrow-line regions where parsec-scale jets (radio cores) are absent (e.g., Arp\,187 \citep{ichi16}). In fact, all radio-quiet AGN are candidates for remnant RLAGN. I refer to radio-quiet AGN here as those sources that are not classified as AGN based on their radio morphology or luminosity, such as for example their optical or infrared spectra or colours. In reality, it is likely that the only difference between these two categories is their jet power, with the radio-loud sources (hosted by significantly more massive galaxies~\citep{best05,best07}) being capable of driving powerful relativistic jets, as seen in radio maps. I refer the reader to \cite{hard18_nature} for a more detailed discussion. If these optically-identified AGN are instead restarting, this would reveal the population of sources with rapidly intermittent activity, with timescales of $\sim$$10^4$--$10^5$\,yr for the light travel time from the central engine to the narrow-line regions, on the order of 1\,kpc. A similar timescale has been attributed to central dimming of the peculiar source `Hanny's Voorwerp' \citep{lint09}, and implied by X-ray observations of its central region~\citep{scha10}, although recent 150\,MHz data from LOFAR have revealed extended radio lobes, suggesting an even older unrelated episode of activity with a radiative timescale of $\sim$$10^8$\,yr \citep{smit22}. Further uses of optical data consist of understanding the hosts of the population between remnant and restarting sources, which has been discussed in Section~\ref{sect:lowfreq}. One of the most scarce pieces of information in the optical and infrared domain relates to the spectroscopic properties of remnants and restarting sources (although SDSS spectroscopy has been analysed by \citep{jurl21_spectral}), as well as line-velocity measurements on small-scales that might provide insights on gas fuelling, as has been found in a few restarting sources \citep[][]{morg21}.

\section{Models and Simulations}
\label{sect:simulations}
\textls[-18]{In this section, I discuss analytic, semi-analytic, and numerical simulations performed to explain the nature of remnant and restarting sources. For active RLAGN on kpc scales, in general, the basic principle of their evolution is that the radio source expands due to the interplay between the properties of the bipolar jets and those of the external environment (e.g.,~\citep{sche74,blan74,kais97}). The radio lobes (which expand due to their internal pressure and the ram pressure of the jet) are significantly over-pressured with respect to the external environment in the early stages of evolution, and drive a shock into the ambient medium until they expand enough in the later stages that they achieve pressure balance (at least in the transverse direction), at which stage buoyancy forces (or infalling cluster gas \citep{engl19}) that push the lobes further from the central engine become dominant. Recent advances to (semi-)analytical modelling describe more realistic RLAGN environments (e.g., \citep{turn15,hard18}) than early works, finding that the lobe growth is not self-similar in terms of the lobe axial ratio, and are detailed enough to be fitted to observations of the lobes and of their surrounding shocks \citep{hard19,maha20}. Numerical modelling, which is more difficult due to the large spatial dynamic range required and due to the lack of knowledge on the magnetic field structure and particle acceleration on various scales, has been important in understanding more detailed lobe physics as well as in understanding AGN feedback (see \citep{norm82,will85,ciof92,hakr13,bour17,mukh18} and references therein). Below, I describe the current status of models used for restarting and remnant sources.}

One of the important differences between active single-cycle jets and restarting jets is that the environment that the jets propagate through is different (under a simplistic model in which all single-cycle jets have not had episodic activity, which can only be challenged in the case of infinite radio sensitivity of the RLAGN population, significant strides can be made to detect remnant lobes in currently known active sources using $n$Jy sensitivity expected in future wide-area radio surveys, as mentioned in Section \ref{sect:nextgen}). As mentioned previously, a restarting jet drives through an environment consisting of the remnant lobe from previous activity (except in cases where the outer lobes are thought to be detached, such as in Cen\,A), which is significantly less dense than the \textit{external} hot gas environments of RLAGN unless the remnant lobes have completely thermalized and mixed with their environment. It is to be expected that multiple and cumulative episodes along the same jet axis would enhance the synchrotron luminosity and size of the relativistic material \citep{chri73}. In this scenario, early numerical simulations \citep{clar91}, in contrast to models of single-cycle sources, predict that the restarted jet is \textit{heavy} with respect to its environment, that is, the remnant lobe. One of the important consequences of this result is that the restarted jet always propagates faster than the previous jet \citep{walg14}; for typical jet velocities, this would mean that the restarted jet requires only a fraction (decreasing with the number of episodes) of the duty cycle to reach the outer edge of the remnant lobe (e.g., \citep{clar91}), which is consistent with the rarity of DDRGs. Thus, radio sources with short intermittency timescales will be larger and brighter than those that spend longer in their remnant phase. Another important property is that due to the nature of the relativistic particles in the remnant lobe the sound speed of the surrounding medium of the restarting jet is greater than that of the hot gas external medium, and the bow shock due to the restarting jet is weaker than the original jet. While these bow shocks are predicted to be readily observable \citep{clar97}, it is unclear why such features are scarcely found in either radio or X-ray observations. However, a few dynamical models predict that the synchrotron emission from the bow shock dominates that from the lobes for a remnant at GHz frequencies, and may be detectable by the future SKA (\citep{ito15}, and see Section \ref{sect:nextgen}). Simulations (i.e., \citep{clar91,clar97}) predict that restarting jets should be well collimated, with only weak termination shocks and dim hotspots, as the external environment of restarting jets consists of remnant lobe plasma, which is expected to be much less dense than the ISM, IGM, or ICM. Another interesting feature of this model is that bow shocks within the remnant plasma driven by the inner lobes should be visible via radio through synchrotron radiation, as the remnant lobes will contain (perhaps mildly relativistic) charged particles and magnetic fields. Apparent bow shocks in the nearby source (widely believed to be restarting \cite{morg99}) Centaurus\,A have been detected \citep{kraf03,kraf07}. Interestingly, the observed radio jets of the inner double of PKS\,B1545-321 are much larger in transverse width (13\,kpc) compared with the collimated jets of non-restarting powerful sources ($\sim$few kpc, e.g., \citep{brid84}). However, high resolution maps of the inner jets reveal a 3.2\,kpc-width collimated channel of surface brightness decrement along the width of the jets, up to the hotspots. In \cite{safo08}, this channel was interpreted as the Doppler-dimmed central collimated jets, which are pointing away from the observer, as predicted by simulations.

\begin{figure}[H]
\includegraphics[width=8 cm]{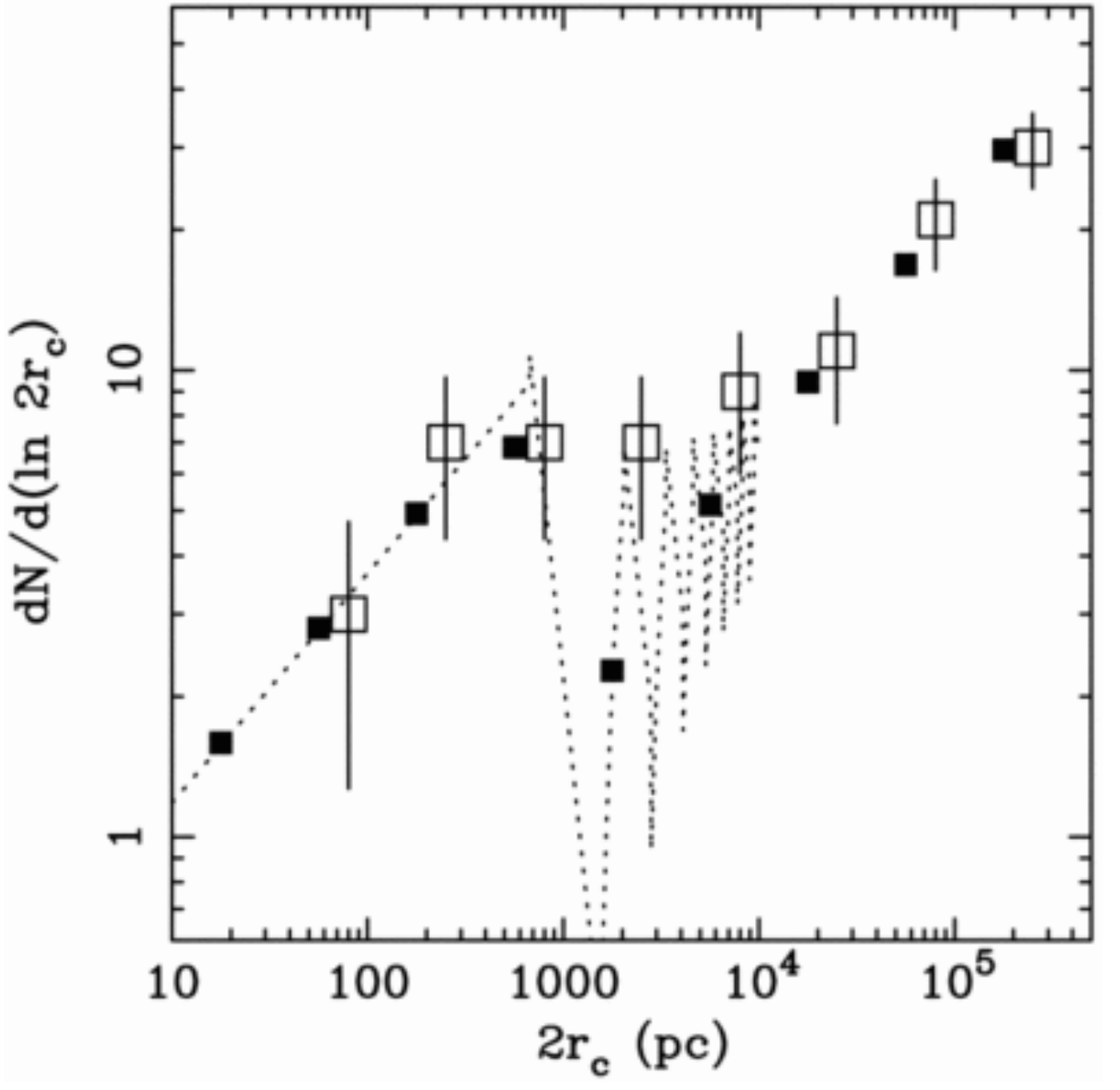}
\caption{Radio source size function from the dynamical models of intermittent radio sources by~\cite{reyn97} (dotted line and binned as black filled squares). Open squares show the data from the CSS/GPS sample by \cite{odea97}, showing the plateau in sizes at $\sim$1 kpc, approximately consistent with theoretical models of intermittent radio sources showing fine structure related to the number of intermittency cycles.}
\label{fig:reynolds_sizedist}
\end{figure}

\textls[-18]{Dynamical models of classical double radio source evolution that have been compared with observed source populations predict the existence of numerous compact and short-lived jets (e.g., \citep{alex00,hard19,shab20}). In general, these types of models aim to determine tracks in the power-linear size (PD) diagram and other distribution functions of sources for a given jet power and environment and to predict whether the radio luminosity of the source will rise with distance (or age) to a peak before radiative losses become dominant and the luminosity decreases on kpc scales and greater. Calibrating these models with observational limitations (e.g., flux limits) generally results in the over-abundance of compact sources, seen observationally as a plateau in the size distribution of radio sources between 100\,pc and 10\,kpc; \citep{odea97}; see Figure~\ref{fig:reynolds_sizedist}. Dynamical modelling by \cite{reyn97} shows that short-intermittency timescales of $10^4$--$10^5$\,yr can explain this over-abundance of small sources. During the remnant phase, their models predict that the shocked shell driving into the external medium remains supersonic, qualitatively explaining the multiple ripples seen in the ICM of Perseus A. In particular, the plateau in the size distribution of sources at small linear sizes can be explained by sources undergoing their first or first few cycles of activity, which in the models is seen as fine structure (periodic evolution) in the size distribution, which is `washed out' observationally by stochasticity (Figure~\ref{fig:reynolds_sizedist}).}

More recent dynamical models of remnant and restarting sources are semi-analytic, using information gained from numerical simulations, and studies have calibrated them using low frequency observations of heterogeneous samples of active, remnant, and restarting sources. In \cite{hard19,shab20}, forward dynamical modelling was used to predict the properties of RLAGN detected by LOFAR at 150\,MHz in the HETDEX and Lockman-Hole fields, respectively. In the former, the study used semi-analytic models \citep{hard18} that were presented for remnant sources to describe their radiative evolution after the jet turns off, predicting that the remnant emission indeed fades rapidly (for more more, see \citep{ito15}). Their prediction for the remnant fraction for an ensemble of modelled sources (which observationally encodes information on the real lifetime distribution of RLAGN) was 37\% for $z\leq1$, quickly dropping to negligible fractions out to $z=2$ when assuming a uniform lifetime distribution (i.e., with sources modelled as having any particular lifetime with equal probability within plausible values), broadly consistent with the modelling of the Lockman-Hole field by \cite{godf17}. In \cite{hard19}, these models were used to infer bulk properties of the $\gtrsim$23,000 sources characterized as RLAGN in LoTSS DR1, finding that a uniform lifetime distribution predicts too few low-luminosity compact sources (which may be young and recently restarted) than actually observed. This over-abundance of compact sources at low luminosities could either represent contamination from star-forming objects or RLAGN that do not satisfy a uniform lifetime distribution that sources $\geqslant$100\,kpc do satisfy. \mbox{In \cite{shab20}}, building on the earlier modelling of the Lockman-Hole field \citep{godf17,brie17}, the authors showed that declining power-law distributions of lifetime and jet power approximately reproduce the fractions of observed active remnant and restarted populations. Therefore, population models with a high fraction of short-lived and low-power jets favour observed samples of both remnant and restarting sources detected at 150\,MHz or general radio samples where compact sources dominate down to low radio luminosities \citep{shab08,hard19}. Thus, the implication is that RLAGN jets are disrupted on a broad range of timescales ($10^4$\,yr to $10^7$\,yr, as found observationally), with a higher likelihood on short timescales ($\leqslant$$10^6$\,yr, supported by the typical lifetimes of GPS/CSS sources). This might suggest different jet (re)triggering mechanisms for different RLAGN populations (e.g., \citep{kavi15,krau19}). It should be noted, however, that analytic models may over-predict remnant fractions for those sources in dense environments where detailed lobe-environment interactions may drive more rapid losses than expected \citep{engl19}.

\begin{figure}[H]
\hspace{-12mm}
\includegraphics[scale=0.9, trim={0 1cm 18cm 3.7cm},clip]{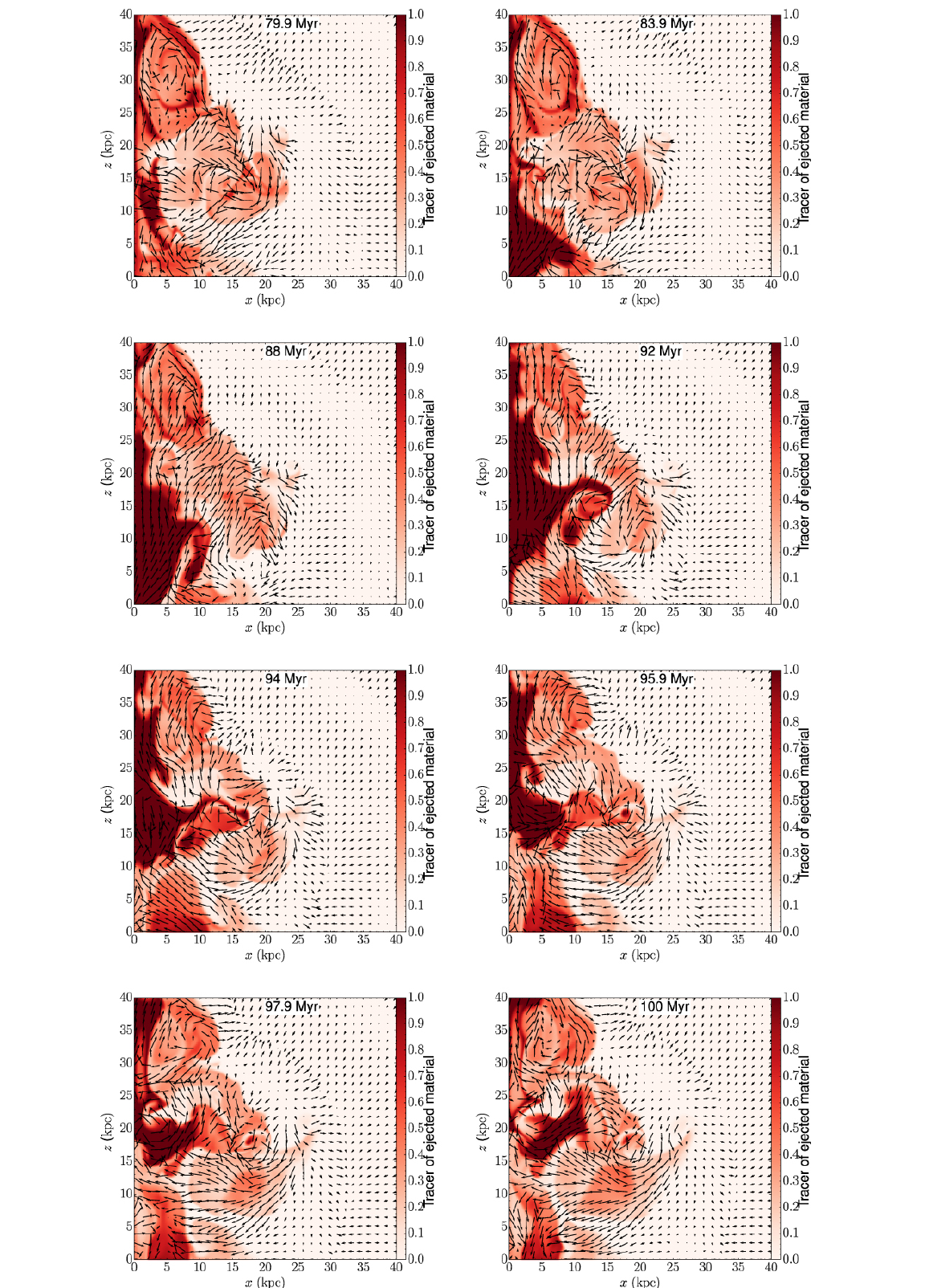}
\caption{Time-stamps (\textbf{left} to \textbf{right}, \textbf{top} to \textbf{bottom}) of the evolution of intermittent RLAGN plasma (taken from \cite{hill17}). The colour-coding is the fraction of material that originated in the jet relative to the environment (dark red = jet plasma, white = ICM). Vectors denote gas velocities. The jet is off in the top left panels, restarting in the top right, turned off in the second row right panel, and remaining as a remnant for the rest. It can be seen that vortices remain intact for the entire remnant phase, and mixing between the jet material and the ICM happens slowly, smoothing out the sporadic nature of intermittency. © AAS, reproduced with permission. }
\label{fig:hillel+17_sim}
\end{figure}

Modelling the effect of the radio source on the environment or the effect of the environment on the radio source (e.g., \citep{mend12}) is key to understanding whether the remnant or restarting phase is energetically important; also key is understanding the expected remnant lobe morphology for sources in different environments, where our current poor understanding drives selection methods of remnant and restarting sources in wide-area surveys (see Section~\ref{sect:lowfreq}). Numerical (magneto-)hydrodynamical simulations prescribe detailed physics to the lobe evolution. It is thought that after the jet turns off, becoming a remnant RLAGN, the bow shock surrounding the remnant lobes continues to expand for a short period; the fraction of input energy in the shock can increase with time during the remnant phase \citep{engl19} until the lobe achieves pressure equilibrium with the surrounding medium. The bow shock eventually vanishes, leaving the radio lobes susceptible to Rayleigh-Taylor instabilities, which cause the lobe and external material to mix over time (e.g., \citep{brug01,hill17}). Through the development of long-lived vortices that survive even in the remnant phase, this mixing is suggested to take place over Myr timescales (\citep{hill17}, and see Figure \ref{fig:hillel+17_sim}) such that the sporadic heating by intermittent jets is smoothed out through a gentle heating process~\citep{mcnm16,hoga17}. This has important consequences, as the environment that the restarting jet drives through determines the energetics of its termination point, the particle energy distribution of the source as a whole, the source detectability, and the overall lobe morphology. In \cite{yate18}, it was found that restarting sources in clusters are significantly more detectable in radio compared with those in poor-group environments. On the other hand, simulations by~\cite{engl19} have shown that, when represented by a $\beta$-model, different cluster parameters have little effect on the growth of the remnant lobe. Galaxy evolution models which include competition between gas cooling and heating by jet-driven shocks (e.g.,~\citep{shab08}) demonstrate that feedback can occur effectively, while intermittent jet activity can occur due to subsequent depletion of the cold gas available for accretion onto the AGN~\citep{pras15}. Large-scale magnetic fields, which are known to pervade the ICM with \textls[-15]{strengths between 1 and 10\,$\upmu$G, can interact with rising remnant lobes; magneto-hydrodynamic} (MHD) simulations~\citep{deyo03,jone05} show that ICM magnetic fields, which are initially dynamically unimportant, can protect the lobes against surface instabilities when amplified through mutual interaction. In addition, it has been suggested that metal enrichment in the ICM is efficient for restarting jets, convecting metals from the host galaxy out to the cluster's peripheries~\citep{huar08}. This provides evidence for co-evolution of remnant lobes and their surrounding medium, as well as for radio lobes remaining intact for timescales $\gtrsim$$10^8$\,yr, and additionally for the idea that feedback effects in clusters are not solely driven by RLAGN in their active state. Radiative models further imply that the spectral evolution of remnants leads to the observed spectra in radio relics (the diffuse cluster sources thought to originate from the lobes of RLAGN \citep{kais02}), and other analytic models imply numerous cavities to be detected by sensitive X-ray instruments \citep{enss02}, implying a seamless transition from RLAGN lobes to large-scale cluster radio sources.

On smaller scales, the extensive observational evidence for restarting jets affecting their surrounding multi-phase ISM on pc--kpc scales (Section \ref{sect:restarting}) is predicted by numerical simulations. In particular, strong coupling is predicted between the small-scale jet and the surrounding clumpy ISM \citep{wagn12,mukh16,mukh18}, reproducing the gas kinematics seen with observations, and they further predict shocked gas being driven into the ISM, as seen on large scales.

The following physical assumptions used in these models and the caveats to them should be borne in mind:
\begin{itemize}
\item We know from observations of remnants and restarting sources that the jet power can drop substantially below radio-detection limits' however, there is no information on whether this drops to zero. Most models assume that the jet turns off completely, but there is no a priori physical reason to assume this. Without information on the power spectrum of jet power variations across the lifetime of a given source, our understanding of their energetics is limited. Hydrodynamical simulations (e.g., \citep{pras15}) have shown results in natural jet power modulation (over a few orders of magnitude in power) related to cycles of accretion activity, and this may be the case in a more substantial way for remnant and restarting sources.
\item Many models include prescriptions for the evolution of classical double-lobed (FR-II) sources, as their dynamics are well-constrained and their models are understood with an observational perspective. Only a few nearby FR-I sources with jets close to the plane of the sky have been robustly modelled (e.g., 3C\,31; \citep{lain02}), but the vast majority of sources at low-redshift are FR-I; thus, the majority of nearby systems that may genuinely be remnant or restarting may be of this type. Moreover, the majority of steep-spectrum sources are found in clusters and are tailed objects, which are difficult to model. Therefore, population models have the major caveat that they are modelled with different jet dynamics than what may actually be true, although the principle nature of jet deceleration on large scales is thought to be the same for both FR-I and FR-II~sources.
\end{itemize}

\section{Summary and Open Questions}
Here, I summarize and discuss the results presented in earlier sections and conclude with prospects for the future from upcoming instruments.
\label{sect:summary}
\subsection{Jet Disruption, Morphology, and Accretion Rate}
Jets from RLAGN are disrupted over a range of lifetimes or length-scales. Even DDRGs, which are all of FR-II type and usually reside in poor environments, tend to have a range of quiescent phases between $10^5$ and $10^7$ yr, and which are at most 50\% of the length of their previous active phase. We know that restarting sources are not special, as they are hosted by galaxies of a similar distribution to the RLAGN population; thus, the evidence points to jet disruption being an inherent process in RLAGN.

Inherent to jet production is the accretion system. Accretion is known to be variable in many objects, but on timescales orders of magnitude lower than the disruption timescales implied by radiative modelling of different episodes of radio activity. There is evidence that selected DDRGs are of the low-excitation (LERG) type in terms of their accretion system (e.g., \citep{maha19}), and in these objects accretion activity by definition cannot be investigated. However, LERGs are believed to be operated by the Chaotic Cold Accretion (CCA \cite{gasp12}) model, a self-regulating process by which cold gas clumps in a multiphase hot gas environment funnel down toward the AGN, promoting episodic spikes in the accretion rate at a level orders of magnitude higher than the traditional Bondi rate and which has been shown to vary on Myr timescales \citep{gasp13}. This is further enhanced by evidence for RLAGN being triggered more frequently in the most massive host systems \citep{shab08}, a higher tendency for restarting jets in LERG FR-IIs than their high-excitation (HERG) counterparts (making for more restarted activity in higher-mass systems \cite{sari12_review}), a tendency for LERGs with cold gas being more radio-loud than LERGs without cold gas \citep{jans12}, the commonality between dust lanes and high molecular gas fractions in various restarting sources (Cen\,A, 3C\,293, 4C\,29.30), and evidence for triggering by accretion of small local gas clouds (e.g., PKS\,B1718-649 \citep{macc14}). Other restarting sources, where the jet disruption timescales are shorter (suggested for more compact objects), could reflect a fluctuating magnetic field structure that dominates jet production \citep{siko13,cham21}; however, this is less well understood observationally. CSS and GPS sources, many of which are restarting, tend to have high accretion rates \citep{wu09}, the conditions of which may favour fluctuating magnetic fields near the accretion system. A systematic selection of a large sample of restarting sources with detailed information on their hot gas environments (which is difficult to obtain even on their own) is required for a more robust conclusion. What will likely remain unknown, however, is whether the jet becomes radio-quiet or radio-silent during the remnant phase.

Do jets always restart along the same jet axis? While this is obviously true for prototypical DDRGs, as they are selected this way, it is unclear whether the rest of the restarted jet population significantly change direction. In \cite{gopa22_ddrg}, DDRG status was associated with the radio source in Abell\,980, which has an inner double and a large-scale outer double significantly offset from the current jet axis; the authors suggest that the outer double has buoyantly risen away from the centre of the ICM. X-shaped sources are the opposite of DDRGs, in the sense that the active jets drive at a large angle from the axis of the more diffuse and older lobe plasma. It is not known whether X-shaped and `winged' sources are truly restarting, as competing models exist for their formation (e.g., large deflection of their back-flow \cite{sari09,josh19,giri22,lala22}) and as X-shaped sources of the LERG type (which are common in DDRGs) are rare. It should be noted that the formation of X-shaped lobes and restarted activity are not necessarily mutually exclusive, as the source PKS\,2014-55 is known to have X-shaped lobes while being a DDRG \citep{cott20}; thus, it is possible that X-shaped sources are truly a result of deflected backflows rather than restarted jets at a different angle. 3C\,293, the well known and studied restarting source, has a 35 deg projected misalignment between the inner and outer lobes, which has been suggested to result either from jet refraction due to pressure gradients in the circumgalactic medium \citep{brid81} or from a change in the black hole spin axis \citep{mach16}. The latter argument is supported by the fact that the radio source is hosted by a post-merger galaxy UGC\,8782, which has two optical nuclei and multiple dust lanes~\citep{cape00}, providing strong evidence for jet axis changes induced by mergers. Thus, it is plausible to suggest that restarted sources as a population are likely to have jets driving along a similar axis to their previous state and that the large deviation in jet axis is due to the buoyant rise of the remnant lobe along steep pressure gradients in the external hot gas~medium.

\subsection{Role in Galaxy and Cluster Evolution}
It is well known that RLAGN, in their active state, have the means to affect galaxy and cluster evolution through shock-heating and various other processes that contribute to the feedback cycle. The key unknown is whether the remnant or restarting state is energetically important. What is the fate of remnant lobe plasma? In \cite{bege01}, it was suggested that the heating effects continue to occur after the radio source has faded while accumulating during the restarted phase. This is supported by simulations of long-lived vortices between the jet material and the external gas that prolong the heating during the remnant phase. The remnant lobes rise buoyantly to the cluster peripheries, suggesting distributed heating and mixing. Even without this mixing, jets are known to restart very quickly after the remnant phase, meaning that shock-driven heating may not cease for long. Low-frequency observations are beginning to observe older remnant plasma associated with cavities that have gone undetected in the past, making it plausible that every massive galaxy or cluster hosts a shock-driving intermittent radio source (e.g., \citep{saba19}). Compact CSO objects which are restarting tend to exist in cool-core clusters (e.g., 3C\,317;  \citep{vent04}); thus, even in environments in which cooling is taking place, RLAGN with rapid intermittency and the potential for cumulative heating episodes may exist. B2\,0258+35 is a restarted source with the new jets driving molecular gas outflows in the ISM, showing that feedback on ISM scales can occur in these systems \citep{murt22}. Many restarted sources have evidence suggesting driving of these fast outflows (e.g., \citep{morg05}), enriching the ISM, with the intermittency driving effects on the kinematics of gas phases on all scales. Along with the various other heating mechanisms that apply to RLAGN in general, observations and simulations suggest that AGN feedback requires episodic activity.

\subsection{Conclusions}
The advances made to date in the study of remnant and restarting sources have shed light on their nature. Below, I summarize the main conclusions following from the studies discussed in this review:
\begin{itemize}
\item Restarted activity, or jet disruption in general, occurs on a variety of timescales ranging from tens of kilo-years to Myr.
\item Restarted and remnant activity is not exclusive to radio-loud objects or to host galaxies of a particular type. Jet disruption is likely stochastic and related to the accretion system.
\item Jets are more likely to restart along the same axis as the previous outburst{, while having similar jet powers between episodes, so that the central engine processes that control jet launching are not significantly changed.}
\item The remnant and restarting phase remains energetically important on all {physical} scales, and {these phases }can be required in order for the AGN feedback cycle to be effective in offsetting cooling {in and around galaxies.}

\end{itemize}

\subsection{Next-Generation Radio Telescopes}
\label{sect:nextgen}
The improvement in depth, resolution, and sky area offered by upcoming and next-\linebreak generation radio telescopes is most rewarding to the study of RLAGN lifecycles. Because remnant and restarting fractions provide constraints on their lifetime distributions, and as many open questions remain, more robust source statistics continue to be required. In particular, deep, sensitive, and complete number counts as a function of the redshift of remnant and restarting sources is expected to be possible in the future. One aspect that has not been discussed in this review is that of radio luminosity functions; embedded in the number density of RLAGN as a function of radio luminosity and redshift is their cosmic evolution (e.g., \citep{cond19}), which can be compared for different types of RLAGN. Indeed, using such information, it has become well known that LERGs and HERGs (i.e., RLAGN with different accretion types) have different cosmic evolutions \citep{best05,prac16,mira17}, and this is associated with the similar conclusion that RLAGN and AGN have different triggering mechanisms~\citep{best05} and different radial density distributions \citep{mach01}. Radio luminosity functions require very deep (e.g., $\upmu$Jy-level) source counts over a wide sky area in order to be as complete as possible in flux and redshift; for this reason alone, luminosity functions for remnant and restarting sources are missing. However, these are expected to become available with the sensitivity levels expected of next-generation instruments (see below), shedding light on the triggering mechanisms, evolution, and relationship with radio-quiet AGN in general. High spatial dynamic range allowing both small and large angular scales to be mapped will facilitate more robust spectral aging studies on the population level, although the caveats discussed in relation to lobe magnetic fields may persist. This is particularly important because jet activity seems to cease at any stage of RLAGN growth, from compact to giant sources, and because their luminosities range over similar orders of magnitude. Raw sensitivity at low frequencies is important as well; the radio source 3C\,452 was known to be a prototypical FR-II source for several decades until deep GMRT observations at 325\,MHz revealed outer remnant structures, and it is now classed as a DDRG \citep{siro13}. Therefore, the notion that every RLAGN has had a previous cycle of activity is comprehensible, and although significant strides have been made with currently new facilities such as LOFAR, uGMRT, MWA, and MeerKAT, the next generation radio telescopes should shed further light on this.

\begin{figure}[H]
\includegraphics[scale=0.4]{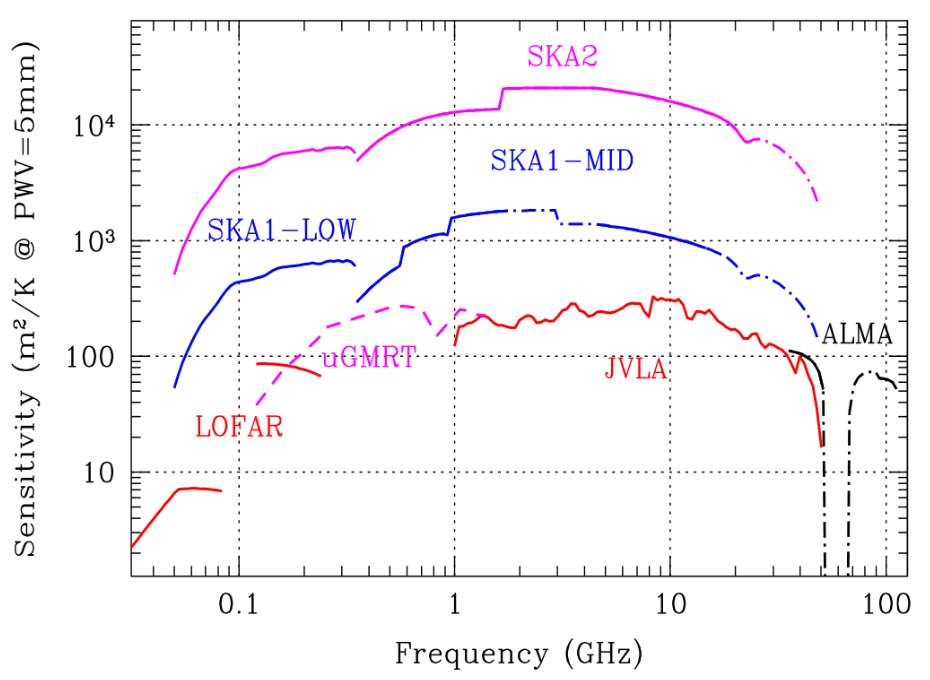}
\caption{Anticipated sensitivity of the SKA1 and SKA2 telescopes as a function of frequency, compared with existing telescopes. Figure taken from 
\url{https://www.skao.int/en/science-users/}, {accessed on 1 February 2023}. Credit: SKA Observatory.} 
\label{fig:skasensitivity}
\end{figure}

\begin{figure}[H]
\includegraphics[scale=0.8, trim={0 1cm 0 12cm},clip]{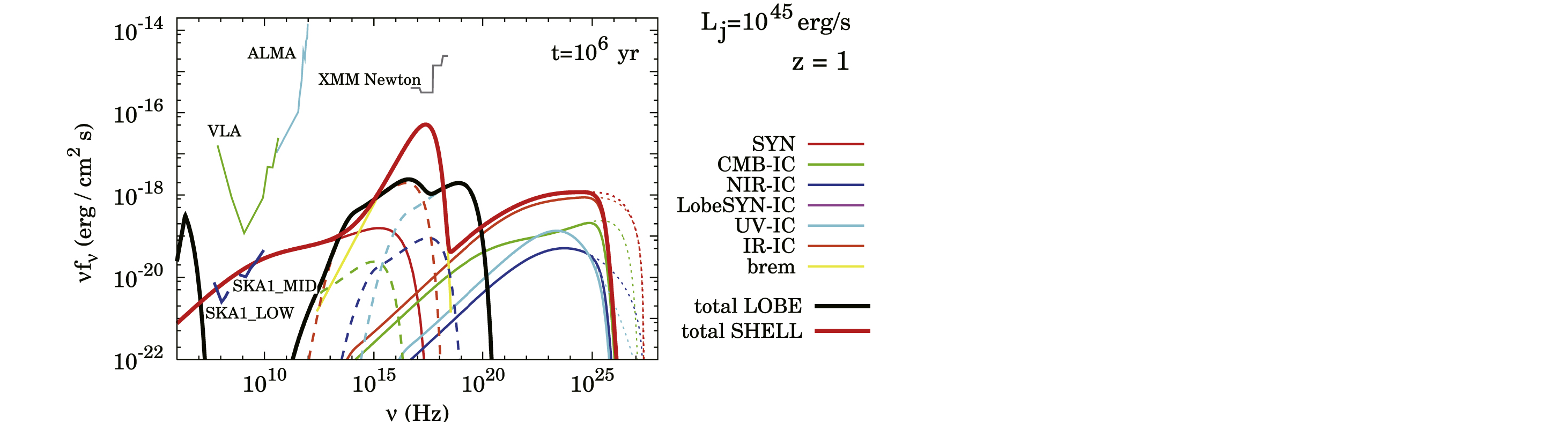}
\caption{{Modelled} 
spectral energy distribution of a typical remnant at $z=1$ and with a jet power of $L_j=10^{45}$\,erg\,s$^{-1}$, taken from \cite{ito15}. They predict that while the total flux from the remnant lobes (thick black line) may not be visible at GHz frequencies with existing telescopes, the bow shock (thick dark red line) may be marginally visible with both SKA1$-$LOW and SKA1$-$MID for an integration time of 10 h. }
\label{fig:ito+15_remnant_sed}
\end{figure}

The SKA1-LOW all-sky survey, which is the low frequency aspect of the upcoming and initial SKA survey \citep{wate16} operating at 50--350\,MHz, is expected to reach RMS noise levels of 20\,$\upmu$Jy at an angular resolution of $\lesssim$10 arcsec, a factor of 4--5 deeper than LoTSS DR2 at a similar frequency, although at a slightly worse angular resolution (note that these parameters define the planned first phase of the SKA design; SKA2, the second phase of the SKA project, is planned to deliver sensitivities and angular resolutions at least a factor of few better). Figure \ref{fig:skasensitivity} shows a comparison of the raw sensitivity expected against that of current telescopes that are used for remnant and restarting source observations at various frequencies. This will enable new detections of remnant plasma in the nearby universe as well as detection of all RLAGN with luminosity above $\sim$$6\times10^{23}\text{\,W\,Hz}^{-1}$ at $z=2$ and of many out to $z\sim5$ (see Figure \ref{fig:murg+05} by \citep{kapi15}). The limitation in angular resolution will make optical identifications of new radio detections difficult, however, and would require follow-up observations of detected objects using currently existing instruments with longer baselines in order to understand their energetics (i.e., if they have a radio core, such as restarting sources; clearly, this would not help for genuine remnants). SKA1-MID (350--24,000 MHz) with arcsec angular resolution will be beneficial for optical IDs as well as in detecting remnant and restarting sources in dense environments in the case that their radiative losses are limited; a remnant source with a luminosity of only $6\times10^{23}\text{\,W\,Hz}^{-1}$ at 160\,MHz with a low frequency spectral index of $\alpha=1.8$ will be detected with SKA1-MID out to $z\approx0.2$, increasing to $z\approx1.6$ if its luminosity is $6\times10^{24}\text{\,W\,Hz}^{-1}$; \citep{kapi15}. Particularly when the full SKA project is complete, this will clearly allow studies of remnant and restarting sources as a function of redshift, which are currently lacking. Moreover, radiative models predict that the observationally-elusive bow shocks that surround remnant plasma (discussed in Section \ref{sect:simulations}) will be detectable with both SKA1-LOW and SKA1-MID \citep{ito15}, as shown in Figure \ref{fig:ito+15_remnant_sed}. This will facilitate a new class of study on `ghost bow shocks' that do not have detectable radio emission from the remnant lobes that drove the bow shocks, and may shed more light on the energetic dominance of RLAGN in feedback even without an active jet.

With the tremendous progress in remnant and restarting source studies driven by developments using LOFAR (Section \ref{sect:lowfreq}), we can expect further progress in the near future with its upgrade, LOFAR 2.0. LOFAR 2.0 is an upgrade to LOFAR's computing power and, being a software telescope, this translates to increased observing capabilities. Specifically, \textls[-5]{observing efficiency is increased, as well as increased bandwidth and simultaneous observing} \textls[-10]{with both its Low Band Antenna (LBA; $\sim$50\,MHz) and High Band Antenna (HBA; $\sim$150\,MHz)} arrays. Coupled with the addition of further stations in Europe, an increase in sensitivity and resolution can be expected, along with rapid surveying time even beyond that which LOFAR already has. This will greatly improve number counts of the faintest radio sources (remnants), and will allow very low frequency spectral information (10--200\,MHz) at an angular resolution ($\gtrsim$1 arcsec at 50\,MHz) that will not be matched by SKA-LOW.

\textls[-10]{Other planned instruments include higher radio frequencies, such as the next generation} VLA (ngVLA), as well as optical and X-ray instruments such as the Vera-Rubin Telescope and Athena, all of which will greatly improve sensitivity and allow new parameter spaces to be probed in the study of remnant and restarting sources.

\vspace{6pt}




\funding{This research received no external funding.}


\dataavailability{Not applicable.}


\conflictsofinterest{The author declares no conflict of interest.}



\abbreviations{Abbreviations}{
The following abbreviations are used in this manuscript:\\

\noindent
\begin{tabular}{@{}ll}
AGN & Active Galactic Nuclei\\
RLAGN & Radio-Loud AGN\\
ISM & InterStellar Medium\\
IGM & InterGalactic Medium\\
ICM & IntraCluster Medium\\
DDRG & Double-Double Radio Galaxy \\
CSS & Compact Steep Spectrum\\
\end{tabular}

\noindent
\begin{tabular}{@{}ll}

GPS & Gigahertz Steep Spectrum\\
VLA & Very Large Array\\
LOFAR & LOw Frequency ARray\\
GMRT & Giant Metrewave Radio Telescope\\
MWA & Murchinson Wide-field Array\\
SKA & Square Kilometer Array\\
\end{tabular}
}


\begin{adjustwidth}{-\extralength}{0cm}

\reftitle{References}

\PublishersNote{}
\end{adjustwidth}
\end{document}